\documentclass[aps,pre,preprint,groupedaddress]{revtex4-1}

\usepackage{mathtools}
\usepackage{amsmath}
\usepackage{amssymb}
\usepackage[english]{babel}

\usepackage{color}

\newcommand{\SIText}{Appendix}

\begin{document}

\title{Energy transfer from large to small scales in turbulence by multi-scale nonlinear strain and vorticity interactions}

\author{Perry L. Johnson}
\email{perryj@stanford.edu}

\affiliation{Center for Turbulence Research, Stanford University, Stanford, CA 94305, USA}

\date{\today}

\begin{abstract}
An intrinsic feature of turbulent flows is an enhanced rate of mixing and kinetic energy dissipation due to the rapid generation of small-scale motions from large-scale excitation. The transfer of kinetic energy from large to small scales is commonly attributed to the stretching of vorticity by the strain-rate, but {\color{black} strain self-amplification also plays a role. Previous treatments of this connection are phenomenological or inexact, or cannot distinguish the contribution of vorticity stretching from that of strain self-amplification.} In this paper, an exact relationship is derived which {\color{black} quantitatively} establishes how intuitive multi-scale mechanisms such as vorticity stretching and strain self-amplification together actuate the inter-scale transfer of energy in turbulence. Numerical evidence {\color{black} validates this result and uses it to} demonstrate that the contribution of strain self-amplification to energy transfer is higher than that of vorticity stretching, but not overwhelmingly so.
\end{abstract}


\maketitle

\section{Introduction}
Fluid turbulence is an archetypal nonlinear multi-scale phenomenon in classical physics. Encounters with turbulent flows are ubiquitous in both the natural sciences and engineering, due to the small viscosities of common fluids like air and water relative to the typical sizes and velocities in many flows. Turbulent flows are generally characterized by a continuous spectrum of energetic length and time scales, and understanding how these scales dynamically interact is a cardinal matter for turbulence modeling. The ability of turbulence to quickly produce small scale motions from large scale excitation has traditionally been characterized as a `cascade' of energy, which has become a linchpin for the study of turbulence physics \cite{Richardson1922, Kolmogorov1941, Onsager1949, Frisch1995, Falkovich2009, Biferale2003}.

The stretching of vorticity by the strain-rate has been traditionally viewed as the basic mechanism by which energy is transferred from large to small scales \cite{Taylor1938, Onsager1949, Pullin1998}. In this view, coherent regions of high rotation rate (or vorticity) are preferentially subjected to extensional flow (positive strain-rate) along the axis of rotation. The conservation of angular momentum requires an increase in vorticity magnitude accompanied by a decrease in cross-section. The result is positive work done by the strain-rate on the vortex resulting in activity at smaller length scales \cite{Tennekes1972}. This concept of vortex stretching has been very influential and many studies of inter-scale energy transfer in turbulence have focused on it \cite{Lundgren1982, Jimenez1998, Chorin1988, Lozano2016, Doan2018}.

A statistical {\color{black}(or global)} connection {\color{black}has been established between the net amplfication of vorticity by the strain-rate and the net energy transfer to small scales using the Karman-Howarth equations} \cite{Karman1938}. While the analogy to material line stretching \cite{Taylor1938} is not perfect because vorticity does not have the same alignment behavior as passive material lines \cite{Holzner2010, Johnson2016, Johnson2017a}, the vorticity preferentially aligns with the strain-rate eigenvector having the second largest eigenvalue, which tends to be extensional \cite{Vieillefosse1982, Vieillefosse1984, Ashurst1987, Cantwell1992}. 

The statistical connection between vorticity stretching and the energy cascade is not unique, however. An equally valid candidate mechanism is strain-rate self-amplification, i.e., the steepening of compressive strain-rates via nonlinear self-advection \cite{Tsinober2009}. The positive average vorticity stretching cannot be disentangled from positive average strain self-amplification in homogeneous, or approximately locally homogeneous, flows \cite{Betchov1956}. Furthermore, truncated series expansions suggest that strain self-amplification contributes three times more than vorticity stretching to inter-scale energy transfer \cite{Eyink2006, Carbone2019}. 

{\color{black}The notion of spectral blocking in two-dimensional turbulence \cite{Fjortoft1953} due to the conservation of enstrophy highlights in a more precise qualitative way that vorticity stretching (which vanishes in 2D) is necessary for sustained energy transfer toward small scales. However, strain self-amplification also vanishes in 2D, and the same line of reasoning applied to the dissipation rate demonstrates that strain self-amplification is simultaneously necessary. Thus, this approach cannot distinguish between the contribution of vorticity stretching or strain self-amplification to the energy `cascade'.}

Explanations of vorticity stretching often invoke different length scales of organized strain-rate and vorticity, but (unfiltered) velocity gradients emphasize dynamics at the smallest scales \cite{Meneveau2011}. Spatially filtered velocity gradients are more suited to describe behavior in the inertial range where the energy `cascade' is a dominant feature \cite{Danish2018}. Previous approaches using spatial filtering and/or velocity increments \cite{Borue1998, Eyink2006, Carbone2019} have connected inertial range inter-scale energy transfer with vorticity stretching and strain self-amplification, but have essentially done so by truncating an infinite series, which leaves uncertainty regarding the role of neglected higher-order terms.

\textit{In this paper, an exact connection is demonstrated between inter-scale energy transfer, i.e., the `energy cascade', and spatio-temporally localized multi-scale interactions of vorticity and strain-rate in a turbulent flow.} {\color{black}The derived relationship is validated using direct numerical simulations, and then it is further leveraged to} reveal the true extent to which vorticity stretching and strain self-amplification at various scales contribute to the transfer of energy from large to small scales.


\section{Derivation}
The velocity field, $\mathbf{u}(\mathbf{x}, t)$, of an incompressible turbulent flow evolves according to,
\begin{equation}
\frac{\partial u_i}{\partial t} + u_j \frac{\partial u_i}{\partial x_j} = - \frac{1}{\rho} \frac{\partial p}{\partial x_i} + \nu \nabla^2 u_i + f_i,
\label{eq:Navier-Stokes}
\end{equation}
where $\rho$ is the fluid mass density, $\nu$ is the kinematic viscosity of the fluid, and $\mathbf{f}$ is any forcing applied to the fluid. The pressure field, $p(\mathbf{x}, t)$, enforces the divergence-free constraint, $\nabla \cdot \mathbf{u} = 0$. The velocity gradient tensor, $A_{ij} = \partial u_i / \partial x_j$, describes the local flow topology in terms of strain-rate, $S_{ij} = \tfrac{1}{2}\left(A_{ij} + A_{ji}\right)$ and rotation-rate, $\Omega_{ij} = \tfrac{1}{2}\left(A_{ij} - A_{ji}\right)$, which can also be expressed as the vorticity vector, $\omega_{i} = \epsilon_{ijk} \Omega_{kj}$.

A turbulent flow with mean kinetic energy $\langle K \rangle = \tfrac{1}{2} \langle u_i u_i \rangle$ and mean dissipation rate $\langle \epsilon \rangle = 2\nu \langle S_{ij} S_{ij} \rangle$ is characterized by a wide range of length scales from an integral length scale, $L \sim \langle K \rangle^{3/2} \langle \epsilon \rangle^{-1}$, down to the Kolmogorov length scale, $\eta = \nu^{3/4} \langle \epsilon \rangle^{-1/4}$. The dynamic range of a turbulent flow increases as $L / \eta \sim Re_\lambda^{3/2}$, where $Re_\lambda \sim \langle K \rangle / \sqrt{\nu \langle \epsilon \rangle}$ is the Taylor-scale Reynolds number.

The features of a turbulent velocity field larger than a given scale $\ell$ can be isolated using a low-pass filter \cite{Germano1992},
\begin{equation}
\overline{u}_i^{\ell} = G_\ell \star u_i, ~~~ \mathcal{F}\{\overline{u}_i^{\ell}\} = \mathcal{F}\{G_\ell\} \mathcal{F}\{u_i\},
\label{eq:generic-filter}
\end{equation}
where $\mathcal{F}\{\cdot\}$ denotes the Fourier transform and $\star$ denotes the convolution operator. The superscript, $\ell$ in this case, denotes the filter width. 
The evolution equation for the large-scale dynamics is obtained by filtering Eq.\ \eqref{eq:Navier-Stokes},
\begin{equation}
\frac{\partial \overline{u}_i^\ell}{\partial t} + \overline{u}_j^\ell \frac{\partial \overline{u}_i^\ell}{\partial x_j} = - \frac{1}{\rho} \frac{\partial \overline{p}^\ell}{\partial x_i} + \nu \nabla^2 \overline{u}_i^\ell + \overline{f}_i^\ell - \frac{\partial \sigma_{ij}^\ell}{\partial x_j},
\end{equation}
where $\sigma_{ij}^\ell = \overline{u_i u_j}^\ell - \overline{u}_i^\ell \overline{u}_j^\ell$ represents an `effective stress' on the large-scale velocity caused by features smaller than $\ell$. The kinetic energy at scales larger than $\ell$ is defined as $E^{\ell}(\mathbf{x}, t) = \tfrac{1}{2} \overline{\mathbf{u}}_i^{\ell} \overline{\mathbf{u}}_i^{\ell} $, and $e^{\ell}(\mathbf{x}, t) = \tfrac{1}{2}\sigma_{ii}^\ell$ represents the kinetic energy at scales smaller than $\ell$. The large- and small-scale energies evolve according to,
\begin{gather}
\frac{\partial E^\ell}{\partial t} + \frac{\partial T_{i}^\ell}{\partial x_i} = \overline{u}_i^\ell \overline{f}_i^\ell - \Pi^\ell - \mathcal{E}^\ell,
\label{eq:large-scale-energy}\\
\frac{\partial e^\ell}{\partial t} + \frac{\partial t_{i}^\ell}{\partial x_i} = q^\ell + \Pi^\ell - \varepsilon^\ell
\label{eq:small-scale-energy}
\end{gather}
where $T_{i}^\ell$ and $t_{i}^\ell$ describe spatial redistribution of large- and small-scale energy, respectively (see \cite{Germano1992} for more details). The molecular dissipation rate of large- and small-scale energy is $\mathcal{E}^\ell = 2 \nu \overline{S}_{ij}^\ell \overline{S}_{ij}^\ell$ and $\varepsilon^\ell = 2 \nu ( \overline{S_{ij} S_{ij}}^\ell - \overline{S}_{ij}^\ell \overline{S}_{ij}^\ell )$, respectively. The work done by forcing on the small scales is $q^\ell = \overline{u_i f_i} - \overline{u}_i \overline{f}_i$. The term $\Pi^\ell = -\sigma_{ij}^\ell \overline{S}_{ij}^\ell$ appears in these two equations with opposite sign, representing energy transfer between large and small scales across scale $\ell$. If energy is injected by forcing at large scales, then for a(n) (approximately) steady homogeneous flow with $\eta \ll \ell \ll L$, the energy balance becomes,
\begin{equation}
\langle u_i f_i \rangle \approx \langle \overline{u}_i^\ell \overline{f}_i^\ell \rangle \approx \langle \Pi^\ell \rangle \approx \langle \varepsilon^\ell \rangle \approx \langle \epsilon \rangle.
\label{eq:inertial-range}
\end{equation}
For the present purposes, the validity of Eq.\ \eqref{eq:inertial-range} defines the inertial range of scales, where the exchange of energy across $\ell$ by $\Pi^\ell$ is from large to small scales in the mean in order to facilitate the dissipation of kinetic energy predominantly at small scales.


\begin{figure}[tbp]
	\centering
	\includegraphics[trim = 13mm 15mm 17mm 5mm, clip, width=0.49\linewidth]{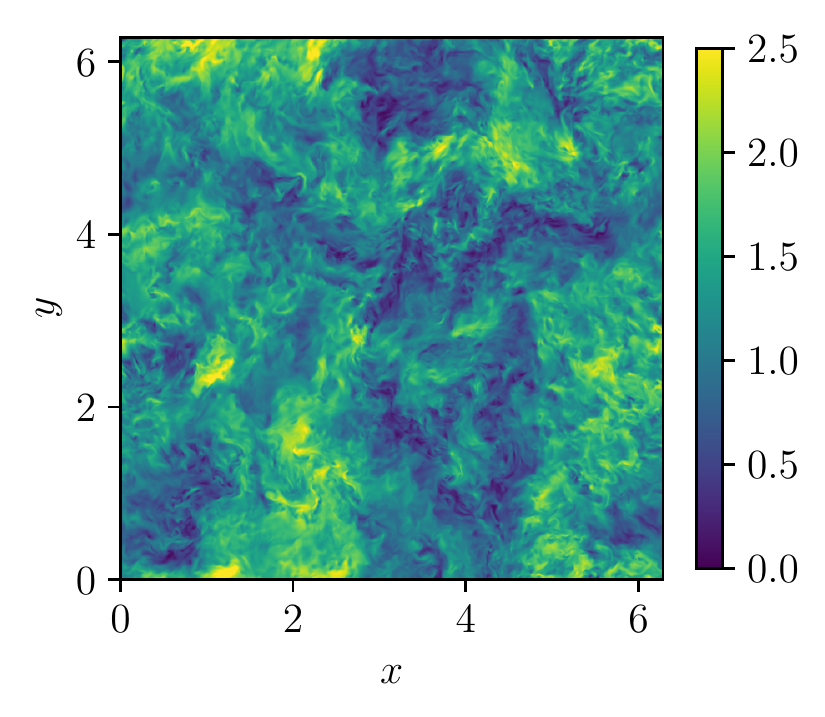}
	\includegraphics[trim = 13mm 15mm 17mm 5mm, clip, width=0.49\linewidth]{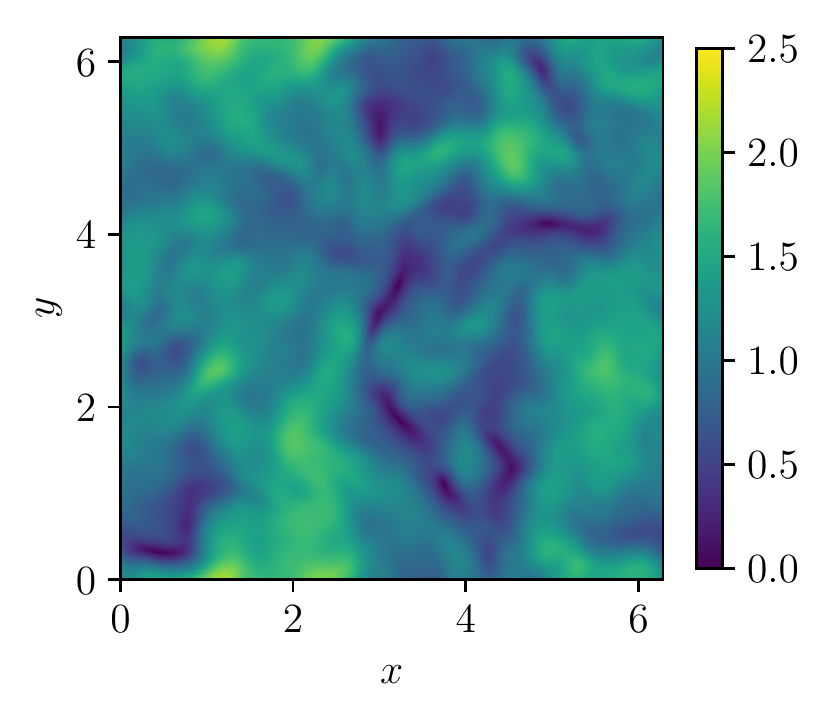}
	\caption{Unfiltered (left) and filtered (right) velocity magnitude on a slice through the 3D forced isotropic turbulence simulation at $Re_\lambda = 400$. A filter width of $\ell = 35 \eta$ is used.}
	\label{fig:snapshot}
\end{figure}

In the following, a Gaussian low-pass filter,
\begin{equation}
\begin{gathered}
G_{\ell}(\mathbf{r}) =  \mathcal{N} e^{-|\mathbf{r}|^2 / (2 \ell^2)}, ~~~ \mathcal{F}\{G_{\ell}\}(\mathbf{k}) = e^{-|\mathbf{k}|^2 \ell^2/2},
\end{gathered}
\label{eq:Gaussian-filter}
\end{equation}
with $\mathcal{N} = \left( 2 \pi \ell^2 \right)^{-3/2}$, is used to derive a spatio-temporally local relationship between filtered velocity gradients and the transfer flux of energy across $\ell$ from large to small scales. 
Figure \ref{fig:snapshot} shows velocity magnitude on a slice in a turbulent flow before and after the application of a Gaussian filter. It may be readily seen from Eqs.\ \eqref{eq:generic-filter} and \eqref{eq:Gaussian-filter} that $\overline{\mathbf{u}}^\ell$ is the solution of the diffusion equation,
\begin{equation}
\frac{\partial \overline{u}_i^\ell}{\partial (\ell^2)} = \frac{1}{2} \nabla^2 \overline{u}_i^\ell, ~~~ \overline{u}_i^{\ell=0} = u_i(\mathbf{x}, t),
\label{eq:parabolic-filter}
\end{equation}
where $\ell^2$ is the time-like variable. Using the definition of $\sigma_{ij}$ with Eq.\ \eqref{eq:parabolic-filter}, it is straightforward to show that the effective sub-filter scale stress may be obtained as a solution of a forced diffusion equation,
\begin{equation}
\frac{\partial \sigma_{ij}^{\ell}}{\partial (\ell^2)} = \frac{1}{2} \nabla^2 \sigma_{ij}^{\ell} + \overline{A}_{ik}^{\ell} \overline{A}_{jk}^{\ell}, ~~~ \sigma_{ij}^{\ell=0} = 0,
\label{eq:parabolic-stress}
\end{equation}
where $\overline{A}_{ij}^\ell$ is the filtered velocity gradient tensor.

The solution to Eq. \eqref{eq:parabolic-stress}, with the Gaussian kernel as the Green's function, and can be written as,
\begin{equation}
\sigma_{ij}^\ell = \int_{0}^{\ell^2} d\theta \left( \overline{\overline{A}_{ik}^{\sqrt{\theta}} \overline{A}_{jk}^{\sqrt{\theta}}}^{\sqrt{\ell^2-\theta}} \right).
\label{eq:stress}
\end{equation}
In this way, the sub-filter stress is the collective result of contributions from velocity gradient fields filtered at all scales $\sqrt{\theta}$ smaller than $\ell$. The filter at $\sqrt{\ell^2 - \theta}$ projects these contributions onto the larger scales.

The integrand of Eq.\ \eqref{eq:stress} bears some resemblance to the nonlinear model \cite{Clark1979, Borue1998}, $\sigma_{ij}^\ell \approx \ell^2 \overline{A}_{ik}^\ell \overline{A}_{jk}^\ell$, but differs from such previous expressions in that Eq.\ \eqref{eq:stress} is exact rather than an approximate relation formed by truncating an infinite series. Furthermore, Eq.\ \eqref{eq:stress} straightforwardly decomposes into scale-local and scale-nonlocal components,
\begin{equation}
\sigma_{ij}^\ell = \ell^2 \overline{A}_{ik}^{\ell} \overline{A}_{jk}^{\ell} 
+ \int_{0}^{\ell^2} d\theta \left( \overline{\overline{A}_{ik}^{\sqrt{\theta}} \overline{A}_{jk}^{\sqrt{\theta}}}^{\phi} - \overline{\overline{A}_{ik}^{\sqrt{\theta}}}^{\phi} \overline{\overline{A}_{jk}^{\sqrt{\theta}}}^{\phi} \right),
\label{eq:stress-split}
\end{equation}
where $\phi = \sqrt{\ell^2 - \theta}$. The first term on the right side of Eq. \eqref{eq:stress-split} is `scale-local' because it only involves quantities resolved at scale $\ell$. The second term involves the difference of the filtered product and the product of filtered quantities, representing the contributions of sub-filter scale velocity gradients to the stress. This is considered `scale-nonlocal' because it contains velocity gradients at finer scales than $\ell$. More specifically, `locality' in this context is referring to ultraviolet locality \cite{Eyink2005}.

Contracting Eq.\ \eqref{eq:stress-split} with the filtered strain-rate tensor forms an expression for $\Pi^\ell = - \sigma_{ij}^\ell \overline{S}_{ij}^\ell$. Then, substituting the decomposition $A_{ij} = S_{ij} + \Omega_{ij}$ leads to
\begin{equation}
\begin{gathered}
\Pi^\ell = \Pi_{l,S}^\ell + \Pi_{l,\Omega}^\ell + \Pi_{nl,S}^\ell + \Pi_{nl,\Omega}^\ell + \Pi_{nl,c}^\ell, \\
\text{where} \\
\Pi_{l,S}^{\ell} = -\ell^2 \overline{S}_{ij}^{\ell} \overline{S}_{jk}^{\ell} \overline{S}_{ki}^{\ell}, \hspace{0.1\linewidth}
\Pi_{l,\Omega}^{\ell} = \tfrac{1}{4}\ell^2 \overline{\omega}_{i}^{\ell} \overline{S}_{ij}^{\ell} \overline{\omega}_{j}^{\ell},
\\
\Pi_{nl,S}^{\ell} = -\int_{0}^{\ell^2} d\theta \left( \overline{\overline{S}_{ik}^{\sqrt{\theta}} \overline{S}_{jk}^{\sqrt{\theta}}}^{\phi} - \overline{\overline{S}_{ik}^{\sqrt{\theta}}}^{\phi} \overline{\overline{S}_{jk}^{\sqrt{\theta}}}^{\phi} \right)\overline{S}_{ij}^{\ell},
\\
\Pi_{nl,\Omega}^{\ell} = \frac{1}{4} \int_{0}^{\ell^2} d\theta \left( \overline{\overline{\omega}_{i}^{\sqrt{\theta}} \overline{\omega}_{j}^{\sqrt{\theta}}}^{\phi} - \overline{\overline{\omega}_{i}^{\sqrt{\theta}}}^{\phi} \overline{\overline{\omega}_{j}^{\sqrt{\theta}}}^{\phi} \right)\overline{S}_{ij}^{\ell},
\\
\Pi_{nl,c}^{\ell} = \int_{0}^{\ell^2} d\theta \left( \overline{\overline{S}_{ik}^{\sqrt{\theta}} \overline{\Omega}_{jk}^{\sqrt{\theta}}}^{\phi} + \overline{\overline{\Omega}_{ik}^{\sqrt{\theta}} \overline{S}_{jk}^{\sqrt{\theta}}}^{\phi} \right)\overline{S}_{ij}^{\ell}.
\end{gathered}
\label{eq:Pi-decomposition}
\end{equation}
The first two terms in \eqref{eq:Pi-decomposition} represent inter-scale energy transfer by scale-local strain-self amplification ($\Pi_{l,S}$) and scale-local vorticity stretching ($\Pi_{l,\Omega}$), respectively. By themselves, these two terms comprise the nonlinear model of Ref.\ \cite{Clark1979} and are given the subscript `l' to denote `scale-local', expressing the fact that these terms involve only quantities filtered at scale $\ell$. The remaining three terms have the subscript `nl' for `nonlocal', indicating that these quantities involve smaller scales than $\ell$. {\color{black} These `nonlocal' terms include interactions of scales only slightly smaller than $\ell$, so a more intricate discussion of `cascade' locality is included in the \SIText.} The third and fourth terms represent the amplification by strain at scale $\ell$ of sub-filter strain ($\Pi_{nl,S}$) and sub-filter vorticity ($\Pi_{nl,\Omega}$). The fifth term represents energy transfer by the resolved strain-rate tensor acting on the sub-filter correlation of strain-rate and vorticity. This energy exchange mechanism is less intuitive and has not received much attention, with the exception of \cite{Eyink2006}.


\textit{The decomposition, \eqref{eq:Pi-decomposition}, is exact and establishes a direct relationship, at a particular location and time in a flow, between the energy flux across scale $\ell$ and the multi-scale interaction of vorticity and strain.}
This result enables the systematic decomposition of turbulent inter-scale energy transfer in terms of multi-scale interactions such as vorticity stretching and strain self-amplification.

\section{Simulation Results}
To leverage this result, direct numerical simulations of steady homogeneous isotropic turbulence were performed using Eq.\ \eqref{eq:Navier-Stokes} in a triply-periodic box with forcing $\mathbf{f}$ specified such that the energy in the first two wavenumber shells remains constant. Results for a simulation with $Re_\lambda = 400$ having $1024^3$ points in each direction are shown here. The range of active length scales is $L / \eta = 460$. 
Figure \ref{fig:snapshot} illustrates the numerical simulation and filtering procedures. 

\begin{figure}[tbp]
	\centering
	\includegraphics[trim = 7mm 0mm 0mm 0mm, clip, width=0.5\linewidth]{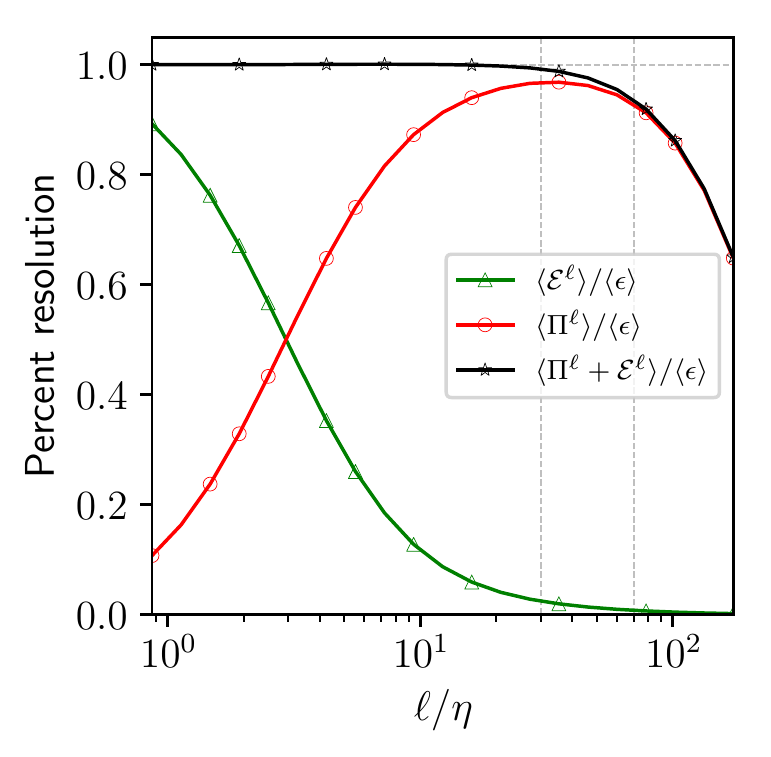}
	\caption{The resolved dissipation rate and the net inter-scale energy transfer as a function of scale using a Gaussian filter on forced isotropic turbulence at $Re_\lambda = 400$. The vertical dashed gray lines indicate $\ell = 30 \eta$ and $\ell = 70 \eta = 0.15 L$.}
	\label{fig:resolution}
\end{figure}

The main features of energy transfer and dissipation in the simulation are shown in Figure \ref{fig:resolution} as a function of filter width, $\ell$. For increasing filter width above $\eta$, the resolved dissipation rate, $\mathcal{E}^\ell$, decreases sharply and most of the viscous energy dissipation is unresolved for $\ell \gg \eta$. On the other hand,  the sum of $\mathcal{E}^\ell$ and $\Pi^\ell$ is equal to the total dissipation rate provided $\ell \ll L$, which indicates the the forcing $\mathbf{f}$ is relatively inactive at these scales, see Eqs.\ \eqref{eq:large-scale-energy} and \eqref{eq:small-scale-energy}. Thus, for a range of scales, $\eta \ll \ell \ll L$, the net energy transfer is equal to the total dissipation rate and Eq.\ \eqref{eq:inertial-range} is approximately satisfied.

\begin{figure}[tbp]
	\centering
	\includegraphics[width=0.6\linewidth]{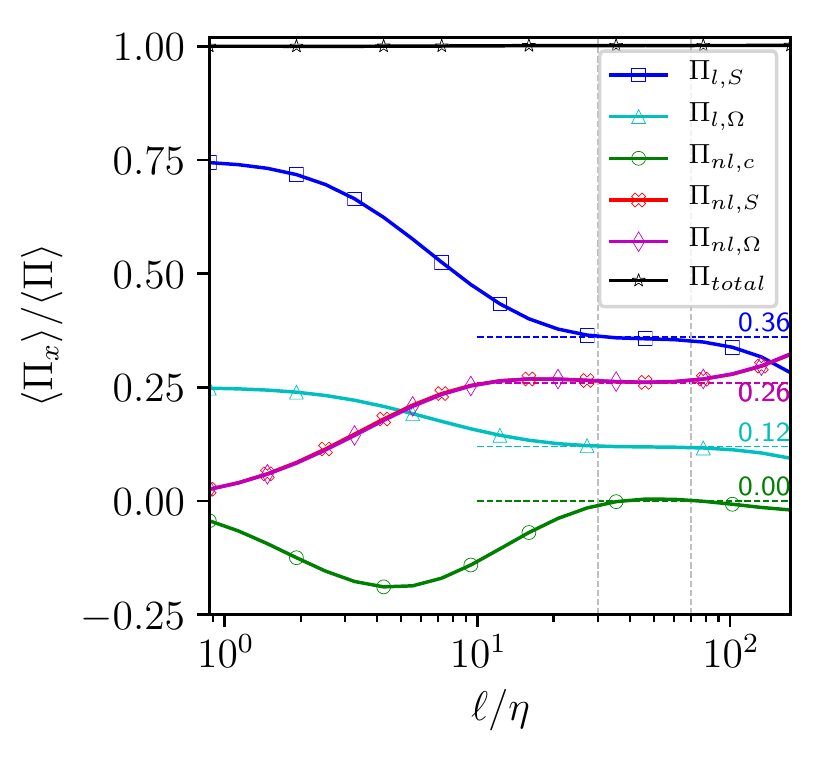}
	\caption{The fraction of net energy transfer, $\langle \Pi^\ell \rangle$, accomplished by the five mechanism from Eq.\ \eqref{eq:Pi-decomposition}. The horizontal dashed lines are added manually to highlight the range of scales for which the composition of inter-scale energy transfer is approximately constant. The vertical dashed gray lines indicate $\ell = 30 \eta$ and $\ell = 70 \eta = 0.15 L$.}
	\label{fig:net-transfer}
\end{figure}

Figure \ref{fig:net-transfer} shows the net contribution from each of the five terms in Eq.\ \eqref{eq:Pi-decomposition} as a function of filter size. The integrals are evaluated using the trapezoidal rule with a discretization over logarithmically distributed points in scale-space ($\theta$) from $0.75 \eta^2$ to $\ell^2$ using roughly $15$ points per decade. {\color{black}First, it is important to point out that the derived relation, Eq.\ \eqref{eq:Pi-decomposition}, is validated by the black line marked with star symbols indicating} $\langle \Pi_{total}^\ell \rangle / \langle \Pi^\ell \rangle = 1$. {\color{black}In other words, this confirms verifies that the ratio of the right and left sides of Eq.\ \eqref{eq:Pi-decomposition} is exactly unity for all filter widths.} {\color{black}Next, consider separately each of the five terms on the right side of Eq.\ \eqref{eq:Pi-decomposition}.} For $\ell \lesssim \eta$, the nonlocal terms are small and the two local terms dominate. The Betchov relation, $\langle \Pi_{l,S}^\ell \rangle = 3 \langle \Pi_{l,\Omega}^\ell \rangle$, constrains the ratio of the two local terms for any $\ell$ in homogeneous incompressible flows \cite{Betchov1956, Carbone2019}. As a consequence, scale-local strain self-amplification is responsible for three times more net energy transfer than scale-local vorticity stretching at any filter width. 

{\color{black} For a range of scales approximately bounded by vertical dashed gray lines in Figures \ref{fig:resolution} and \ref{fig:net-transfer}}, the proportional contribution of each term in Eq.\ \eqref{eq:Pi-decomposition} remains fairly constant in this range of filter widths. The results show that roughly half of the net inter-scale energy transfer in the inertial range is accounted for by the local terms $\Pi_{l,S}^\ell$ and $\Pi_{l,\Omega}^\ell$. The other half is contributed by their nonlocal counterparts, $\Pi_{nl,S}^\ell$ and $\Pi_{nl,\Omega}^\ell$. In contrast to the scale-local terms, the scale-nonlocal terms indicate an even division between strain amplification and vorticity stretching on average. 
Due to the `pirouette' effect \cite{Xu2011}, vorticity is known to align more efficiently with larger-scale, slower evolving strain-rates than with the strain-rate at the same scale \cite{Leung2012, Fiscaletti2016}. With more efficient vorticity stretching, the net inter-scale energy transfer by scale-nonlocal interactions is more evenly balanced between the two mechanisms.

The net contribution of the cross term, $\Pi_{nl,c}^\ell$, is negligible in the inertial range, but provides net backscatter at smaller scales, possibly related to the bottleneck phenomenon \cite{Falkovich1994}. 
This reveals an interesting similarity between 2D and 3D turbulence. Of the five constituents in Eq. \eqref{eq:Pi-decomposition}, only $\Pi_{nl,c}^\ell$ is non-zero in 2D turbulence due to geometric constraints. It is well established that 2D turbulence exhibits net backscatter \cite{Chen2006} associated with an `inverse cascade' of energy \cite{Falkovich2009}, with important consequences for, e.g., rotating turbulence \cite{Buzzicotti2018}.

To summarize, the fractional contributions of net inter-scale energy transfer from each of the five mechanisms in the inertial range can be approximately summarized as $\langle \Pi_{l,S}^\ell \rangle : \langle \Pi_{l,\Omega}^\ell \rangle : \langle \Pi_{nl,S}^\ell \rangle : \langle \Pi_{nl,\Omega}^\ell \rangle : \langle \Pi_{nl,c}^\ell \rangle \approx 3 : 1 : 2 : 2 : 0$. Including scale-local and nonlocal terms together, the ratio of contributions from strain self-amplification and vorticity stretching is roughly $\langle \Pi_{S}^\ell \rangle : \langle \Pi_{\Omega}^\ell \rangle \approx 5:3$. This result stands in contrast to both the traditional view which focuses only on vorticity stretching, as well as more recent views that strain self-amplification is the dominant mechanism, including the view that over-emphasizes that $\langle \Pi_{l,S}^\ell \rangle : \langle \Pi_{l,\Omega}^\ell \rangle = 3:1$ due to the Betchov relation \cite{Carbone2019}. {\color{black}The precise values found for these relative contributions are reported in Figure \ref{fig:net-transfer} are not emphasized because of the limited extent of inertial range provided by the present simulation. Reynolds number effects are further explored in the \SIText, and future work at higher Reynolds numbers can refine these results.} Dependence on filter shape is addressed in the \SIText, and it is expected that the main conclusions will hold for other filter shapes.



\section{Conclusion}
In conclusion, an exact relationship between inter-scale energy transfer and multi-scale vorticity-strain interactions is introduced {\color{black} and validated}. {\color{black}This development disentangles the respective impacts of vorticity stretching and strain self-amplification on the energy `cascade'.} Analysis of detailed simulations reveals that, while scale-local strain self-amplification provides three times the energy transfer as scale-local vorticity stretching, it is just as important to consider multi-scale interactions. For scale-nonlocal interactions, in fact, the net contribution by vorticity stretching and strain self-amplification is roughly equal. As a result, strain self-amplification is responsible for more net inter-scale energy transfer than vorticity stretching, but not overwhelmingly so. Both processes seem important in the rapid production of small-scale motions in turbulence. 

The present view of the inter-scale energy transfer will facilitate a more detailed exploration of the energy cascade in turbulence. For instance, the efficiency of the cascade is known to be quite low \cite{Ballouz2018}, and the present results provide a framework for future exploration of how the cascade is driven by multi-scale velocity gradient dynamics \cite{Xu2011, Fiscaletti2016}. In fact, the present work suggests that it may be more advantageous to pursue shell models expressed in terms of velocity gradients \cite{Biferale2007, Johnson2017b}. Also, the results shown here have focused on the net energy transfer, but this quantity fluctuates in space and time. 
Analysis of fluctuations and negative transfer events, as well as investigations connecting the present work with spatially coherent structures \cite{Bermejo2008, Dong2019}, may also provide a deeper mechanistic understanding of turbulent dynamics. The approach outlined here can be extended to flows with additional physics such as stratification, rotation, chemical reactions, multiple phases, and active matter. 

The insights from this approach provide guidance for advancing models for large-eddy simulations, which are designed to provide accurate results despite under-resolution of turbulent flows on coarse numerical grids \cite{Meneveau2000, Sagaut2006}. The stretching of sub-filter vorticity is an appealing basis for models \cite{Pullin1994, Misra1997, Matheou2014, Silvis2017}, but the analysis here reveals a path for improving on such an approach.


\section*{AcknowledgEments}
The author would like to acknowledge support from the Advanced Simulation and Computing program of the US Department of Energy's National Nuclear Security Administration via the PSAAP-II Center at Stanford, Grant No.\ DE-NA0002373. The author thanks Theo Drivas, as well as Adrian Lozano-Duran and Parviz Moin for fruitful discussions on the topic. 

\bibliographystyle{apsrev4-1}
\bibliography{interscale_energy_transfer.bib}

\begin{thebibliography}{58}%
\makeatletter
\providecommand \@ifxundefined [1]{%
 \@ifx{#1\undefined}
}%
\providecommand \@ifnum [1]{%
 \ifnum #1\expandafter \@firstoftwo
 \else \expandafter \@secondoftwo
 \fi
}%
\providecommand \@ifx [1]{%
 \ifx #1\expandafter \@firstoftwo
 \else \expandafter \@secondoftwo
 \fi
}%
\providecommand \natexlab [1]{#1}%
\providecommand \enquote  [1]{``#1''}%
\providecommand \bibnamefont  [1]{#1}%
\providecommand \bibfnamefont [1]{#1}%
\providecommand \citenamefont [1]{#1}%
\providecommand \href@noop [0]{\@secondoftwo}%
\providecommand \href [0]{\begingroup \@sanitize@url \@href}%
\providecommand \@href[1]{\@@startlink{#1}\@@href}%
\providecommand \@@href[1]{\endgroup#1\@@endlink}%
\providecommand \@sanitize@url [0]{\catcode `\\12\catcode `\$12\catcode
  `\&12\catcode `\#12\catcode `\^12\catcode `\_12\catcode `\%12\relax}%
\providecommand \@@startlink[1]{}%
\providecommand \@@endlink[0]{}%
\providecommand \url  [0]{\begingroup\@sanitize@url \@url }%
\providecommand \@url [1]{\endgroup\@href {#1}{\urlprefix }}%
\providecommand \urlprefix  [0]{URL }%
\providecommand \Eprint [0]{\href }%
\providecommand \doibase [0]{http://dx.doi.org/}%
\providecommand \selectlanguage [0]{\@gobble}%
\providecommand \bibinfo  [0]{\@secondoftwo}%
\providecommand \bibfield  [0]{\@secondoftwo}%
\providecommand \translation [1]{[#1]}%
\providecommand \BibitemOpen [0]{}%
\providecommand \bibitemStop [0]{}%
\providecommand \bibitemNoStop [0]{.\EOS\space}%
\providecommand \EOS [0]{\spacefactor3000\relax}%
\providecommand \BibitemShut  [1]{\csname bibitem#1\endcsname}%
\let\auto@bib@innerbib\@empty
\bibitem [{\citenamefont {Richardson}(1922)}]{Richardson1922}%
  \BibitemOpen
  \bibfield  {author} {\bibinfo {author} {\bibfnamefont {L.~F.}\ \bibnamefont
  {Richardson}},\ }\href@noop {} {\emph {\bibinfo {title} {Weather Prediction
  by Numerical Process}}}\ (\bibinfo  {publisher} {Cambridge},\ \bibinfo {year}
  {1922})\BibitemShut {NoStop}%
\bibitem [{\citenamefont {Kolmogorov}(1941)}]{Kolmogorov1941}%
  \BibitemOpen
  \bibfield  {author} {\bibinfo {author} {\bibfnamefont {A.~N.}\ \bibnamefont
  {Kolmogorov}},\ }\href@noop {} {\bibfield  {journal} {\bibinfo  {journal}
  {Dokl. Akad. Nauk SSSR}\ }\textbf {\bibinfo {volume} {30}},\ \bibinfo {pages}
  {299} (\bibinfo {year} {1941})}\BibitemShut {NoStop}%
\bibitem [{\citenamefont {Onsager}(1949)}]{Onsager1949}%
  \BibitemOpen
  \bibfield  {author} {\bibinfo {author} {\bibfnamefont {L.}~\bibnamefont
  {Onsager}},\ }\href@noop {} {\bibfield  {journal} {\bibinfo  {journal} {L.
  Nuovo Cim.}\ }\textbf {\bibinfo {volume} {6}},\ \bibinfo {pages} {279}
  (\bibinfo {year} {1949})}\BibitemShut {NoStop}%
\bibitem [{\citenamefont {Frisch}(1995)}]{Frisch1995}%
  \BibitemOpen
  \bibfield  {author} {\bibinfo {author} {\bibfnamefont {U.}~\bibnamefont
  {Frisch}},\ }\href@noop {} {\emph {\bibinfo {title} {Turbulence}}}\ (\bibinfo
   {publisher} {Cambridge},\ \bibinfo {year} {1995})\BibitemShut {NoStop}%
\bibitem [{\citenamefont {Falkovich}(2009)}]{Falkovich2009}%
  \BibitemOpen
  \bibfield  {author} {\bibinfo {author} {\bibfnamefont {G.}~\bibnamefont
  {Falkovich}},\ }\href@noop {} {\bibfield  {journal} {\bibinfo  {journal} {J.
  Phys. A-Math. Theor.}\ }\textbf {\bibinfo {volume} {42}},\ \bibinfo {pages}
  {123001} (\bibinfo {year} {2009})}\BibitemShut {NoStop}%
\bibitem [{\citenamefont {Biferale}(2003)}]{Biferale2003}%
  \BibitemOpen
  \bibfield  {author} {\bibinfo {author} {\bibfnamefont {L.}~\bibnamefont
  {Biferale}},\ }\href@noop {} {\bibfield  {journal} {\bibinfo  {journal}
  {Annu. Rev. Fluid Mech.}\ }\textbf {\bibinfo {volume} {35}},\ \bibinfo
  {pages} {441} (\bibinfo {year} {2003})}\BibitemShut {NoStop}%
\bibitem [{\citenamefont {Taylor}(1938)}]{Taylor1938}%
  \BibitemOpen
  \bibfield  {author} {\bibinfo {author} {\bibfnamefont {G.~I.}\ \bibnamefont
  {Taylor}},\ }\href@noop {} {\bibfield  {journal} {\bibinfo  {journal} {P. R.
  Soc. London A}\ }\textbf {\bibinfo {volume} {164}},\ \bibinfo {pages} {15}
  (\bibinfo {year} {1938})}\BibitemShut {NoStop}%
\bibitem [{\citenamefont {Pullin}\ and\ \citenamefont
  {Saffman}(1998)}]{Pullin1998}%
  \BibitemOpen
  \bibfield  {author} {\bibinfo {author} {\bibfnamefont {D.~I.}\ \bibnamefont
  {Pullin}}\ and\ \bibinfo {author} {\bibfnamefont {P.~G.}\ \bibnamefont
  {Saffman}},\ }\href@noop {} {\bibfield  {journal} {\bibinfo  {journal} {Annu.
  Rev. Fluid Mech.}\ }\textbf {\bibinfo {volume} {30}},\ \bibinfo {pages} {31}
  (\bibinfo {year} {1998})}\BibitemShut {NoStop}%
\bibitem [{\citenamefont {Tennekes}\ and\ \citenamefont
  {Lumley}(1972)}]{Tennekes1972}%
  \BibitemOpen
  \bibfield  {author} {\bibinfo {author} {\bibfnamefont {H.}~\bibnamefont
  {Tennekes}}\ and\ \bibinfo {author} {\bibfnamefont {J.~L.}\ \bibnamefont
  {Lumley}},\ }\href@noop {} {\emph {\bibinfo {title} {A First Course in
  Turbulence}}}\ (\bibinfo  {publisher} {MIT Press},\ \bibinfo {year}
  {1972})\BibitemShut {NoStop}%
\bibitem [{\citenamefont {Lundgren}(1982)}]{Lundgren1982}%
  \BibitemOpen
  \bibfield  {author} {\bibinfo {author} {\bibfnamefont {T.~S.}\ \bibnamefont
  {Lundgren}},\ }\href {\doibase 10.1063/1.863957} {\bibfield  {journal}
  {\bibinfo  {journal} {Phys. Fluids}\ }\textbf {\bibinfo {volume} {25}},\
  \bibinfo {pages} {2193} (\bibinfo {year} {1982})}\BibitemShut {NoStop}%
\bibitem [{\citenamefont {Jimenez}\ and\ \citenamefont
  {Wray}(1998)}]{Jimenez1998}%
  \BibitemOpen
  \bibfield  {author} {\bibinfo {author} {\bibfnamefont {J.}~\bibnamefont
  {Jimenez}}\ and\ \bibinfo {author} {\bibfnamefont {A.~A.}\ \bibnamefont
  {Wray}},\ }\href@noop {} {\bibfield  {journal} {\bibinfo  {journal} {J. Fluid
  Mech.}\ }\textbf {\bibinfo {volume} {373}},\ \bibinfo {pages} {255–285}
  (\bibinfo {year} {1998})}\BibitemShut {NoStop}%
\bibitem [{\citenamefont {Chorin}(1988)}]{Chorin1988}%
  \BibitemOpen
  \bibfield  {author} {\bibinfo {author} {\bibfnamefont {A.~J.}\ \bibnamefont
  {Chorin}},\ }\href@noop {} {\bibfield  {journal} {\bibinfo  {journal}
  {Commun. Math. Phys.}\ }\textbf {\bibinfo {volume} {114}},\ \bibinfo {pages}
  {167} (\bibinfo {year} {1988})}\BibitemShut {NoStop}%
\bibitem [{\citenamefont {Lozano-Duran}\ \emph {et~al.}(2016)\citenamefont
  {Lozano-Duran}, \citenamefont {Holzner},\ and\ \citenamefont
  {Jimenez}}]{Lozano2016}%
  \BibitemOpen
  \bibfield  {author} {\bibinfo {author} {\bibfnamefont {A.}~\bibnamefont
  {Lozano-Duran}}, \bibinfo {author} {\bibfnamefont {M.}~\bibnamefont
  {Holzner}}, \ and\ \bibinfo {author} {\bibfnamefont {J.}~\bibnamefont
  {Jimenez}},\ }\href@noop {} {\bibfield  {journal} {\bibinfo  {journal} {J.
  Fluid Mech.}\ }\textbf {\bibinfo {volume} {803}},\ \bibinfo {pages}
  {356–394} (\bibinfo {year} {2016})}\BibitemShut {NoStop}%
\bibitem [{\citenamefont {Doan}\ \emph {et~al.}(2018)\citenamefont {Doan},
  \citenamefont {Swaminathan}, \citenamefont {Davidson},\ and\ \citenamefont
  {Tanahashi}}]{Doan2018}%
  \BibitemOpen
  \bibfield  {author} {\bibinfo {author} {\bibfnamefont {N.~A.}\ \bibnamefont
  {Doan}}, \bibinfo {author} {\bibfnamefont {N.}~\bibnamefont {Swaminathan}},
  \bibinfo {author} {\bibfnamefont {P.~A.}\ \bibnamefont {Davidson}}, \ and\
  \bibinfo {author} {\bibfnamefont {M.}~\bibnamefont {Tanahashi}},\ }\href@noop
  {} {\bibfield  {journal} {\bibinfo  {journal} {Phys. Rev. Fluids}\ }\textbf
  {\bibinfo {volume} {3}},\ \bibinfo {pages} {1} (\bibinfo {year}
  {2018})}\BibitemShut {NoStop}%
\bibitem [{\citenamefont {de~Karman}\ and\ \citenamefont
  {Howarth}(1938)}]{Karman1938}%
  \BibitemOpen
  \bibfield  {author} {\bibinfo {author} {\bibfnamefont {T.}~\bibnamefont
  {de~Karman}}\ and\ \bibinfo {author} {\bibfnamefont {L.}~\bibnamefont
  {Howarth}},\ }\href@noop {} {\bibfield  {journal} {\bibinfo  {journal} {P
  Roy. Soc. A-Math. Phy.}\ }\textbf {\bibinfo {volume} {164}},\ \bibinfo
  {pages} {192} (\bibinfo {year} {1938})}\BibitemShut {NoStop}%
\bibitem [{\citenamefont {Holzner}\ \emph {et~al.}(2010)\citenamefont
  {Holzner}, \citenamefont {Guala}, \citenamefont {Luthi}, \citenamefont
  {Liberzon}, \citenamefont {Nikitin}, \citenamefont {Kinzelbach},\ and\
  \citenamefont {Tsinober}}]{Holzner2010}%
  \BibitemOpen
  \bibfield  {author} {\bibinfo {author} {\bibfnamefont {M.}~\bibnamefont
  {Holzner}}, \bibinfo {author} {\bibfnamefont {M.}~\bibnamefont {Guala}},
  \bibinfo {author} {\bibfnamefont {B.}~\bibnamefont {Luthi}}, \bibinfo
  {author} {\bibfnamefont {A.}~\bibnamefont {Liberzon}}, \bibinfo {author}
  {\bibfnamefont {N.}~\bibnamefont {Nikitin}}, \bibinfo {author} {\bibfnamefont
  {W.}~\bibnamefont {Kinzelbach}}, \ and\ \bibinfo {author} {\bibfnamefont
  {A.}~\bibnamefont {Tsinober}},\ }\href@noop {} {\bibfield  {journal}
  {\bibinfo  {journal} {Phys. Fluids}\ }\textbf {\bibinfo {volume} {22}},\
  \bibinfo {pages} {061701} (\bibinfo {year} {2010})}\BibitemShut {NoStop}%
\bibitem [{\citenamefont {Johnson}\ and\ \citenamefont
  {Meneveau}(2016)}]{Johnson2016}%
  \BibitemOpen
  \bibfield  {author} {\bibinfo {author} {\bibfnamefont {P.~L.}\ \bibnamefont
  {Johnson}}\ and\ \bibinfo {author} {\bibfnamefont {C.}~\bibnamefont
  {Meneveau}},\ }\href@noop {} {\bibfield  {journal} {\bibinfo  {journal}
  {Phys. Rev. E}\ }\textbf {\bibinfo {volume} {93}},\ \bibinfo {pages} {033118}
  (\bibinfo {year} {2016})}\BibitemShut {NoStop}%
\bibitem [{\citenamefont {Johnson}\ \emph {et~al.}(2017)\citenamefont
  {Johnson}, \citenamefont {Hamilton}, \citenamefont {Burns},\ and\
  \citenamefont {Meneveau}}]{Johnson2017a}%
  \BibitemOpen
  \bibfield  {author} {\bibinfo {author} {\bibfnamefont {P.~L.}\ \bibnamefont
  {Johnson}}, \bibinfo {author} {\bibfnamefont {S.~S.}\ \bibnamefont
  {Hamilton}}, \bibinfo {author} {\bibfnamefont {R.}~\bibnamefont {Burns}}, \
  and\ \bibinfo {author} {\bibfnamefont {C.}~\bibnamefont {Meneveau}},\
  }\href@noop {} {\bibfield  {journal} {\bibinfo  {journal} {Phys. Rev.
  Fluids}\ }\textbf {\bibinfo {volume} {2}},\ \bibinfo {pages} {014605}
  (\bibinfo {year} {2017})}\BibitemShut {NoStop}%
\bibitem [{\citenamefont {Vieillefosse}(1982)}]{Vieillefosse1982}%
  \BibitemOpen
  \bibfield  {author} {\bibinfo {author} {\bibfnamefont {P.}~\bibnamefont
  {Vieillefosse}},\ }\href
  {http://books.google.com/books?hl=en{\&}lr={\&}id=n5cE8uCzRwcC{\&}oi=fnd{\&}pg=PA1{\&}dq=Perfect+Incompressible+Fluids{\&}ots=TXYT-Hi3ok{\&}sig=3z6L42MnaHFvfk7uNCi9Z8T12hg}
  {\bibfield  {journal} {\bibinfo  {journal} {J. Phys.-Paris}\ }\textbf
  {\bibinfo {volume} {43}},\ \bibinfo {pages} {837} (\bibinfo {year}
  {1982})}\BibitemShut {NoStop}%
\bibitem [{\citenamefont {Vieillefosse}(1984)}]{Vieillefosse1984}%
  \BibitemOpen
  \bibfield  {author} {\bibinfo {author} {\bibfnamefont {P.}~\bibnamefont
  {Vieillefosse}},\ }\href@noop {} {\bibfield  {journal} {\bibinfo  {journal}
  {Physica A}\ }\textbf {\bibinfo {volume} {125}},\ \bibinfo {pages} {150}
  (\bibinfo {year} {1984})}\BibitemShut {NoStop}%
\bibitem [{\citenamefont {Ashurst}\ \emph {et~al.}(1987)\citenamefont
  {Ashurst}, \citenamefont {Kerstein}, \citenamefont {Kerr},\ and\
  \citenamefont {Gibson}}]{Ashurst1987}%
  \BibitemOpen
  \bibfield  {author} {\bibinfo {author} {\bibfnamefont {W.~T.}\ \bibnamefont
  {Ashurst}}, \bibinfo {author} {\bibfnamefont {A.~R.}\ \bibnamefont
  {Kerstein}}, \bibinfo {author} {\bibfnamefont {R.~M.}\ \bibnamefont {Kerr}},
  \ and\ \bibinfo {author} {\bibfnamefont {C.~H.}\ \bibnamefont {Gibson}},\
  }\href@noop {} {\bibfield  {journal} {\bibinfo  {journal} {Phys. Fluids}\
  }\textbf {\bibinfo {volume} {30}},\ \bibinfo {pages} {2343} (\bibinfo {year}
  {1987})}\BibitemShut {NoStop}%
\bibitem [{\citenamefont {Cantwell}(1992)}]{Cantwell1992}%
  \BibitemOpen
  \bibfield  {author} {\bibinfo {author} {\bibfnamefont {B.~J.}\ \bibnamefont
  {Cantwell}},\ }\href@noop {} {\bibfield  {journal} {\bibinfo  {journal}
  {Phys. Fluids}\ }\textbf {\bibinfo {volume} {4}},\ \bibinfo {pages} {782}
  (\bibinfo {year} {1992})}\BibitemShut {NoStop}%
\bibitem [{\citenamefont {Tsinober}(2009)}]{Tsinober2009}%
  \BibitemOpen
  \bibfield  {author} {\bibinfo {author} {\bibfnamefont {A.}~\bibnamefont
  {Tsinober}},\ }\href@noop {} {\emph {\bibinfo {title} {An Informal Conceptual
  Introduction to Turbulence}}}\ (\bibinfo  {publisher} {Springer},\ \bibinfo
  {year} {2009})\BibitemShut {NoStop}%
\bibitem [{\citenamefont {Betchov}(1956)}]{Betchov1956}%
  \BibitemOpen
  \bibfield  {author} {\bibinfo {author} {\bibfnamefont {R.}~\bibnamefont
  {Betchov}},\ }\href@noop {} {\bibfield  {journal} {\bibinfo  {journal} {J.
  Fluid Mech.}\ }\textbf {\bibinfo {volume} {1}},\ \bibinfo {pages} {497}
  (\bibinfo {year} {1956})}\BibitemShut {NoStop}%
\bibitem [{\citenamefont {Eyink}(2006)}]{Eyink2006}%
  \BibitemOpen
  \bibfield  {author} {\bibinfo {author} {\bibfnamefont {G.~L.}\ \bibnamefont
  {Eyink}},\ }\href@noop {} {\bibfield  {journal} {\bibinfo  {journal} {J.
  Fluid Mech.}\ }\textbf {\bibinfo {volume} {549}},\ \bibinfo {pages} {159}
  (\bibinfo {year} {2006})}\BibitemShut {NoStop}%
\bibitem [{\citenamefont {Carbone}\ and\ \citenamefont
  {Bragg}(2019)}]{Carbone2019}%
  \BibitemOpen
  \bibfield  {author} {\bibinfo {author} {\bibfnamefont {M.}~\bibnamefont
  {Carbone}}\ and\ \bibinfo {author} {\bibfnamefont {A.~D.}\ \bibnamefont
  {Bragg}},\ }\href@noop {} {\bibfield  {journal} {\bibinfo  {journal} {arXiv}\
  } (\bibinfo {year} {2019})},\ \Eprint {http://arxiv.org/abs/1906.07144}
  {1906.07144} \BibitemShut {NoStop}%
\bibitem [{\citenamefont {Fjortoft}(1953)}]{Fjortoft1953}%
  \BibitemOpen
  \bibfield  {author} {\bibinfo {author} {\bibfnamefont {R.}~\bibnamefont
  {Fjortoft}},\ }\href@noop {} {\bibfield  {journal} {\bibinfo  {journal}
  {Tellus}\ }\textbf {\bibinfo {volume} {5}},\ \bibinfo {pages} {225} (\bibinfo
  {year} {1953})}\BibitemShut {NoStop}%
\bibitem [{\citenamefont {Meneveau}(2011)}]{Meneveau2011}%
  \BibitemOpen
  \bibfield  {author} {\bibinfo {author} {\bibfnamefont {C.}~\bibnamefont
  {Meneveau}},\ }\href@noop {} {\bibfield  {journal} {\bibinfo  {journal}
  {Annu. Rev. Fluid Mech.}\ }\textbf {\bibinfo {volume} {43}},\ \bibinfo
  {pages} {219} (\bibinfo {year} {2011})}\BibitemShut {NoStop}%
\bibitem [{\citenamefont {Danish}\ and\ \citenamefont
  {Meneveau}(2018)}]{Danish2018}%
  \BibitemOpen
  \bibfield  {author} {\bibinfo {author} {\bibfnamefont {M.}~\bibnamefont
  {Danish}}\ and\ \bibinfo {author} {\bibfnamefont {C.}~\bibnamefont
  {Meneveau}},\ }\href@noop {} {\bibfield  {journal} {\bibinfo  {journal}
  {Phys. Rev. Fluids}\ }\textbf {\bibinfo {volume} {3}},\ \bibinfo {pages}
  {044604} (\bibinfo {year} {2018})}\BibitemShut {NoStop}%
\bibitem [{\citenamefont {Borue}\ and\ \citenamefont
  {Orszag}(1998)}]{Borue1998}%
  \BibitemOpen
  \bibfield  {author} {\bibinfo {author} {\bibfnamefont {V.}~\bibnamefont
  {Borue}}\ and\ \bibinfo {author} {\bibfnamefont {S.~A.}\ \bibnamefont
  {Orszag}},\ }\href@noop {} {\bibfield  {journal} {\bibinfo  {journal} {J.
  Fluid Mech.}\ }\textbf {\bibinfo {volume} {366}},\ \bibinfo {pages} {1}
  (\bibinfo {year} {1998})}\BibitemShut {NoStop}%
\bibitem [{\citenamefont {Germano}(1992)}]{Germano1992}%
  \BibitemOpen
  \bibfield  {author} {\bibinfo {author} {\bibfnamefont {M.}~\bibnamefont
  {Germano}},\ }\href@noop {} {\bibfield  {journal} {\bibinfo  {journal} {J.
  Fluid Mech.}\ }\textbf {\bibinfo {volume} {238}} (\bibinfo {year}
  {1992})}\BibitemShut {NoStop}%
\bibitem [{\citenamefont {Clark}\ \emph {et~al.}(1979)\citenamefont {Clark},
  \citenamefont {Ferziger},\ and\ \citenamefont {Reynolds}}]{Clark1979}%
  \BibitemOpen
  \bibfield  {author} {\bibinfo {author} {\bibfnamefont {R.~A.}\ \bibnamefont
  {Clark}}, \bibinfo {author} {\bibfnamefont {J.~H.}\ \bibnamefont {Ferziger}},
  \ and\ \bibinfo {author} {\bibfnamefont {W.~C.}\ \bibnamefont {Reynolds}},\
  }\href@noop {} {\bibfield  {journal} {\bibinfo  {journal} {Journal of Fluid
  Mechanics}\ }\textbf {\bibinfo {volume} {91}},\ \bibinfo {pages} {1–16}
  (\bibinfo {year} {1979})}\BibitemShut {NoStop}%
\bibitem [{\citenamefont {Eyink}(2005)}]{Eyink2005}%
  \BibitemOpen
  \bibfield  {author} {\bibinfo {author} {\bibfnamefont {G.~L.}\ \bibnamefont
  {Eyink}},\ }\href@noop {} {\bibfield  {journal} {\bibinfo  {journal} {Physica
  D}\ }\textbf {\bibinfo {volume} {207}},\ \bibinfo {pages} {91} (\bibinfo
  {year} {2005})}\BibitemShut {NoStop}%
\bibitem [{\citenamefont {Xu}\ \emph {et~al.}(2011)\citenamefont {Xu},
  \citenamefont {Pumir},\ and\ \citenamefont {Bodenschatz}}]{Xu2011}%
  \BibitemOpen
  \bibfield  {author} {\bibinfo {author} {\bibfnamefont {H.}~\bibnamefont
  {Xu}}, \bibinfo {author} {\bibfnamefont {A.}~\bibnamefont {Pumir}}, \ and\
  \bibinfo {author} {\bibfnamefont {E.}~\bibnamefont {Bodenschatz}},\
  }\href@noop {} {\bibfield  {journal} {\bibinfo  {journal} {Nat. Phys.}\
  }\textbf {\bibinfo {volume} {7}},\ \bibinfo {pages} {709} (\bibinfo {year}
  {2011})}\BibitemShut {NoStop}%
\bibitem [{\citenamefont {Leung}\ \emph {et~al.}(2012)\citenamefont {Leung},
  \citenamefont {Swaminathan},\ and\ \citenamefont {Davidson}}]{Leung2012}%
  \BibitemOpen
  \bibfield  {author} {\bibinfo {author} {\bibfnamefont {T.}~\bibnamefont
  {Leung}}, \bibinfo {author} {\bibfnamefont {N.}~\bibnamefont {Swaminathan}},
  \ and\ \bibinfo {author} {\bibfnamefont {P.~A.}\ \bibnamefont {Davidson}},\
  }\href@noop {} {\bibfield  {journal} {\bibinfo  {journal} {J. Fluid Mech.}\
  }\textbf {\bibinfo {volume} {710}},\ \bibinfo {pages} {453–481} (\bibinfo
  {year} {2012})}\BibitemShut {NoStop}%
\bibitem [{\citenamefont {Fiscaletti}\ \emph {et~al.}(2016)\citenamefont
  {Fiscaletti}, \citenamefont {Elsinga}, \citenamefont {Attili}, \citenamefont
  {Bisetti},\ and\ \citenamefont {Buxton}}]{Fiscaletti2016}%
  \BibitemOpen
  \bibfield  {author} {\bibinfo {author} {\bibfnamefont {D.}~\bibnamefont
  {Fiscaletti}}, \bibinfo {author} {\bibfnamefont {G.~E.}\ \bibnamefont
  {Elsinga}}, \bibinfo {author} {\bibfnamefont {A.}~\bibnamefont {Attili}},
  \bibinfo {author} {\bibfnamefont {F.}~\bibnamefont {Bisetti}}, \ and\
  \bibinfo {author} {\bibfnamefont {O.~R.~H.}\ \bibnamefont {Buxton}},\
  }\href@noop {} {\bibfield  {journal} {\bibinfo  {journal} {Phys. Rev.
  Fluids}\ }\textbf {\bibinfo {volume} {1}},\ \bibinfo {pages} {064405}
  (\bibinfo {year} {2016})}\BibitemShut {NoStop}%
\bibitem [{\citenamefont {Falkovich}(1994)}]{Falkovich1994}%
  \BibitemOpen
  \bibfield  {author} {\bibinfo {author} {\bibfnamefont {G.}~\bibnamefont
  {Falkovich}},\ }\href@noop {} {\bibfield  {journal} {\bibinfo  {journal}
  {Phys. Fluids}\ }\textbf {\bibinfo {volume} {6}},\ \bibinfo {pages} {1411}
  (\bibinfo {year} {1994})}\BibitemShut {NoStop}%
\bibitem [{\citenamefont {Chen}\ \emph {et~al.}(2006)\citenamefont {Chen},
  \citenamefont {Ecke}, \citenamefont {Eyink}, \citenamefont {Rivera},
  \citenamefont {Wan},\ and\ \citenamefont {Xiao}}]{Chen2006}%
  \BibitemOpen
  \bibfield  {author} {\bibinfo {author} {\bibfnamefont {S.}~\bibnamefont
  {Chen}}, \bibinfo {author} {\bibfnamefont {R.~E.}\ \bibnamefont {Ecke}},
  \bibinfo {author} {\bibfnamefont {G.~L.}\ \bibnamefont {Eyink}}, \bibinfo
  {author} {\bibfnamefont {M.}~\bibnamefont {Rivera}}, \bibinfo {author}
  {\bibfnamefont {M.}~\bibnamefont {Wan}}, \ and\ \bibinfo {author}
  {\bibfnamefont {Z.}~\bibnamefont {Xiao}},\ }\href@noop {} {\bibfield
  {journal} {\bibinfo  {journal} {Phys. Rev. Lett.}\ }\textbf {\bibinfo
  {volume} {96}},\ \bibinfo {pages} {084502} (\bibinfo {year}
  {2006})}\BibitemShut {NoStop}%
\bibitem [{\citenamefont {Buzzicotti}\ \emph {et~al.}(2018)\citenamefont
  {Buzzicotti}, \citenamefont {Aluie}, \citenamefont {Biferale},\ and\
  \citenamefont {Linkmann}}]{Buzzicotti2018}%
  \BibitemOpen
  \bibfield  {author} {\bibinfo {author} {\bibfnamefont {M.}~\bibnamefont
  {Buzzicotti}}, \bibinfo {author} {\bibfnamefont {H.}~\bibnamefont {Aluie}},
  \bibinfo {author} {\bibfnamefont {L.}~\bibnamefont {Biferale}}, \ and\
  \bibinfo {author} {\bibfnamefont {M.}~\bibnamefont {Linkmann}},\ }\href@noop
  {} {\bibfield  {journal} {\bibinfo  {journal} {Phys. Rev. Fluids}\ }\textbf
  {\bibinfo {volume} {3}},\ \bibinfo {pages} {034802} (\bibinfo {year}
  {2018})}\BibitemShut {NoStop}%
\bibitem [{\citenamefont {Ballouz}\ and\ \citenamefont
  {Ouellette}(2018)}]{Ballouz2018}%
  \BibitemOpen
  \bibfield  {author} {\bibinfo {author} {\bibfnamefont {J.~G.}\ \bibnamefont
  {Ballouz}}\ and\ \bibinfo {author} {\bibfnamefont {N.~T.}\ \bibnamefont
  {Ouellette}},\ }\href@noop {} {\bibfield  {journal} {\bibinfo  {journal} {J.
  Fluid Mech.}\ }\textbf {\bibinfo {volume} {835}},\ \bibinfo {pages} {1048}
  (\bibinfo {year} {2018})}\BibitemShut {NoStop}%
\bibitem [{\citenamefont {Biferale}\ \emph {et~al.}(2007)\citenamefont
  {Biferale}, \citenamefont {Chevillard}, \citenamefont {Meneveau},\ and\
  \citenamefont {Toschi}}]{Biferale2007}%
  \BibitemOpen
  \bibfield  {author} {\bibinfo {author} {\bibfnamefont {L.}~\bibnamefont
  {Biferale}}, \bibinfo {author} {\bibfnamefont {L.}~\bibnamefont
  {Chevillard}}, \bibinfo {author} {\bibfnamefont {C.}~\bibnamefont
  {Meneveau}}, \ and\ \bibinfo {author} {\bibfnamefont {F.}~\bibnamefont
  {Toschi}},\ }\href@noop {} {\bibfield  {journal} {\bibinfo  {journal} {Phys.
  Rev. Lett.}\ }\textbf {\bibinfo {volume} {98}},\ \bibinfo {pages} {25}
  (\bibinfo {year} {2007})}\BibitemShut {NoStop}%
\bibitem [{\citenamefont {Johnson}\ and\ \citenamefont
  {Meneveau}(2017)}]{Johnson2017b}%
  \BibitemOpen
  \bibfield  {author} {\bibinfo {author} {\bibfnamefont {P.~L.}\ \bibnamefont
  {Johnson}}\ and\ \bibinfo {author} {\bibfnamefont {C.}~\bibnamefont
  {Meneveau}},\ }\href@noop {} {\bibfield  {journal} {\bibinfo  {journal}
  {Phys. Rev. Fluids}\ }\textbf {\bibinfo {volume} {2}},\ \bibinfo {pages}
  {072601(R)} (\bibinfo {year} {2017})}\BibitemShut {NoStop}%
\bibitem [{\citenamefont {Bermejo-Moreno}\ and\ \citenamefont
  {Pullin}(2008)}]{Bermejo2008}%
  \BibitemOpen
  \bibfield  {author} {\bibinfo {author} {\bibfnamefont {I.}~\bibnamefont
  {Bermejo-Moreno}}\ and\ \bibinfo {author} {\bibfnamefont {D.~I.}\
  \bibnamefont {Pullin}},\ }\href@noop {} {\bibfield  {journal} {\bibinfo
  {journal} {J. Fluid Mech.}\ }\textbf {\bibinfo {volume} {603}},\ \bibinfo
  {pages} {101–135} (\bibinfo {year} {2008})}\BibitemShut {NoStop}%
\bibitem [{\citenamefont {Dong}\ \emph {et~al.}(2019)\citenamefont {Dong},
  \citenamefont {Huang}, \citenamefont {Yuan},\ and\ \citenamefont
  {Lozano-Duran}}]{Dong2019}%
  \BibitemOpen
  \bibfield  {author} {\bibinfo {author} {\bibfnamefont {S.}~\bibnamefont
  {Dong}}, \bibinfo {author} {\bibfnamefont {Y.}~\bibnamefont {Huang}},
  \bibinfo {author} {\bibfnamefont {X.}~\bibnamefont {Yuan}}, \ and\ \bibinfo
  {author} {\bibfnamefont {A.}~\bibnamefont {Lozano-Duran}},\ }\href@noop {}
  {\bibfield  {journal} {\bibinfo  {journal} {arXiv}\ } (\bibinfo {year}
  {2019})}\BibitemShut {NoStop}%
\bibitem [{\citenamefont {Meneveau}\ and\ \citenamefont
  {Katz}(2000)}]{Meneveau2000}%
  \BibitemOpen
  \bibfield  {author} {\bibinfo {author} {\bibfnamefont {C.}~\bibnamefont
  {Meneveau}}\ and\ \bibinfo {author} {\bibfnamefont {J.}~\bibnamefont
  {Katz}},\ }\href@noop {} {\bibfield  {journal} {\bibinfo  {journal} {Annu.
  Rev. Fluid Mech.}\ }\textbf {\bibinfo {volume} {32}},\ \bibinfo {pages} {1}
  (\bibinfo {year} {2000})}\BibitemShut {NoStop}%
\bibitem [{\citenamefont {Sagaut}(2006)}]{Sagaut2006}%
  \BibitemOpen
  \bibfield  {author} {\bibinfo {author} {\bibfnamefont {P.}~\bibnamefont
  {Sagaut}},\ }\href@noop {} {\emph {\bibinfo {title} {Large Eddy Simulation
  for Incompressible Flows}}}\ (\bibinfo  {publisher} {Springer},\ \bibinfo
  {year} {2006})\BibitemShut {NoStop}%
\bibitem [{\citenamefont {Pullin}\ and\ \citenamefont
  {Saffman}(1994)}]{Pullin1994}%
  \BibitemOpen
  \bibfield  {author} {\bibinfo {author} {\bibfnamefont {D.~I.}\ \bibnamefont
  {Pullin}}\ and\ \bibinfo {author} {\bibfnamefont {P.~G.}\ \bibnamefont
  {Saffman}},\ }\href@noop {} {\bibfield  {journal} {\bibinfo  {journal} {Phys.
  Fluids}\ }\textbf {\bibinfo {volume} {6}},\ \bibinfo {pages} {1787} (\bibinfo
  {year} {1994})}\BibitemShut {NoStop}%
\bibitem [{\citenamefont {Misra}\ and\ \citenamefont
  {Pullin}(1997)}]{Misra1997}%
  \BibitemOpen
  \bibfield  {author} {\bibinfo {author} {\bibfnamefont {A.}~\bibnamefont
  {Misra}}\ and\ \bibinfo {author} {\bibfnamefont {D.~I.}\ \bibnamefont
  {Pullin}},\ }\href@noop {} {\bibfield  {journal} {\bibinfo  {journal} {Phys.
  Fluids}\ }\textbf {\bibinfo {volume} {9}},\ \bibinfo {pages} {2443} (\bibinfo
  {year} {1997})}\BibitemShut {NoStop}%
\bibitem [{\citenamefont {Chung}\ and\ \citenamefont
  {Matheou}(2014)}]{Matheou2014}%
  \BibitemOpen
  \bibfield  {author} {\bibinfo {author} {\bibfnamefont {D.}~\bibnamefont
  {Chung}}\ and\ \bibinfo {author} {\bibfnamefont {G.}~\bibnamefont
  {Matheou}},\ }\href@noop {} {\bibfield  {journal} {\bibinfo  {journal} {J.
  Atmos. Sci.}\ }\textbf {\bibinfo {volume} {71}},\ \bibinfo {pages} {1863}
  (\bibinfo {year} {2014})}\BibitemShut {NoStop}%
\bibitem [{\citenamefont {Silvis}\ \emph {et~al.}(2017)\citenamefont {Silvis},
  \citenamefont {Remmerswaal},\ and\ \citenamefont {Verstappen}}]{Silvis2017}%
  \BibitemOpen
  \bibfield  {author} {\bibinfo {author} {\bibfnamefont {M.~H.}\ \bibnamefont
  {Silvis}}, \bibinfo {author} {\bibfnamefont {R.~A.}\ \bibnamefont
  {Remmerswaal}}, \ and\ \bibinfo {author} {\bibfnamefont {R.}~\bibnamefont
  {Verstappen}},\ }\href@noop {} {\bibfield  {journal} {\bibinfo  {journal}
  {Phys. Fluids}\ }\textbf {\bibinfo {volume} {29}},\ \bibinfo {pages} {015105}
  (\bibinfo {year} {2017})}\BibitemShut {NoStop}%
\bibitem [{\citenamefont {Germano}(1986)}]{Germano1986}%
  \BibitemOpen
  \bibfield  {author} {\bibinfo {author} {\bibfnamefont {M.}~\bibnamefont
  {Germano}},\ }\href@noop {} {\bibfield  {journal} {\bibinfo  {journal} {Phys.
  Fluids}\ }\textbf {\bibinfo {volume} {29}},\ \bibinfo {pages} {1755}
  (\bibinfo {year} {1986})}\BibitemShut {NoStop}%
\bibitem [{\citenamefont {{Vreman B}}\ \emph {et~al.}(1994)\citenamefont
  {{Vreman B}}, \citenamefont {{Geurts B}},\ and\ \citenamefont {{Keurten
  H}}}]{Vreman1994b}%
  \BibitemOpen
  \bibfield  {author} {\bibinfo {author} {\bibnamefont {{Vreman B}}}, \bibinfo
  {author} {\bibnamefont {{Geurts B}}}, \ and\ \bibinfo {author} {\bibnamefont
  {{Keurten H}}},\ }\href@noop {} {\bibfield  {journal} {\bibinfo  {journal}
  {J. Fluid Mech.}\ }\textbf {\bibinfo {volume} {278}},\ \bibinfo {pages}
  {p351} (\bibinfo {year} {1994})}\BibitemShut {NoStop}%
\bibitem [{\citenamefont {Lumley}(1992)}]{Lumley1992}%
  \BibitemOpen
  \bibfield  {author} {\bibinfo {author} {\bibfnamefont {J.~L.}\ \bibnamefont
  {Lumley}},\ }\href@noop {} {\bibfield  {journal} {\bibinfo  {journal} {Phys.
  Fluids}\ }\textbf {\bibinfo {volume} {4}},\ \bibinfo {pages} {203} (\bibinfo
  {year} {1992})}\BibitemShut {NoStop}%
\bibitem [{\citenamefont {Aoyama}\ \emph {et~al.}(2005)\citenamefont {Aoyama},
  \citenamefont {Ishihara}, \citenamefont {Kaneda}, \citenamefont {Yokokawa},
  \citenamefont {Itakura},\ and\ \citenamefont {Uno}}]{Aoyama2005}%
  \BibitemOpen
  \bibfield  {author} {\bibinfo {author} {\bibfnamefont {T.}~\bibnamefont
  {Aoyama}}, \bibinfo {author} {\bibfnamefont {T.}~\bibnamefont {Ishihara}},
  \bibinfo {author} {\bibfnamefont {Y.}~\bibnamefont {Kaneda}}, \bibinfo
  {author} {\bibfnamefont {M.}~\bibnamefont {Yokokawa}}, \bibinfo {author}
  {\bibfnamefont {K.}~\bibnamefont {Itakura}}, \ and\ \bibinfo {author}
  {\bibfnamefont {A.}~\bibnamefont {Uno}},\ }\href@noop {} {\bibfield
  {journal} {\bibinfo  {journal} {J. Phys. Soc. Jpn.}\ }\textbf {\bibinfo
  {volume} {74}},\ \bibinfo {pages} {3202} (\bibinfo {year}
  {2005})}\BibitemShut {NoStop}%
\bibitem [{\citenamefont {Domaradzki}\ and\ \citenamefont
  {Carati}(2007)}]{Domaradzki2007}%
  \BibitemOpen
  \bibfield  {author} {\bibinfo {author} {\bibfnamefont {J.~A.}\ \bibnamefont
  {Domaradzki}}\ and\ \bibinfo {author} {\bibfnamefont {D.}~\bibnamefont
  {Carati}},\ }\href@noop {} {\bibfield  {journal} {\bibinfo  {journal} {Phys.
  Fluids}\ }\textbf {\bibinfo {volume} {19}},\ \bibinfo {pages} {085112}
  (\bibinfo {year} {2007})}\BibitemShut {NoStop}%
\bibitem [{\citenamefont {Eyink}\ and\ \citenamefont
  {Aluie}(2009)}]{Eyink2009}%
  \BibitemOpen
  \bibfield  {author} {\bibinfo {author} {\bibfnamefont {G.~L.}\ \bibnamefont
  {Eyink}}\ and\ \bibinfo {author} {\bibfnamefont {H.}~\bibnamefont {Aluie}},\
  }\href@noop {} {\bibfield  {journal} {\bibinfo  {journal} {Phys. Fluids}\
  }\textbf {\bibinfo {volume} {21}},\ \bibinfo {pages} {115107} (\bibinfo
  {year} {2009})}\BibitemShut {NoStop}%
\bibitem [{\citenamefont {Cardesa}\ \emph {et~al.}(2015)\citenamefont
  {Cardesa}, \citenamefont {Vela-Martin}, \citenamefont {Dong},\ and\
  \citenamefont {Jimenez}}]{Cardesa2015}%
  \BibitemOpen
  \bibfield  {author} {\bibinfo {author} {\bibfnamefont {J.~I.}\ \bibnamefont
  {Cardesa}}, \bibinfo {author} {\bibfnamefont {A.}~\bibnamefont
  {Vela-Martin}}, \bibinfo {author} {\bibfnamefont {S.}~\bibnamefont {Dong}}, \
  and\ \bibinfo {author} {\bibfnamefont {J.}~\bibnamefont {Jimenez},
  \bibfnamefont {J.nez}},\ }\href@noop {} {\bibfield  {journal} {\bibinfo
  {journal} {Phys. Fluids}\ }\textbf {\bibinfo {volume} {27}},\ \bibinfo
  {pages} {111702} (\bibinfo {year} {2015})}\BibitemShut {NoStop}%
\bibitem [{\citenamefont {Eyink}(2014)}]{Eyink2014}%
  \BibitemOpen
  \bibfield  {author} {\bibinfo {author} {\bibfnamefont {G.~L.}\ \bibnamefont
  {Eyink}},\ }\href {http://www.ams.jhu.edu/~eyink/TurbulenceIII/notes.html}
  {\enquote {\bibinfo {title} {{Turbulence Theory III}},}\ }\bibinfo
  {howpublished} {course notes, Johns Hopkins University} (\bibinfo {year}
  {2014}),\ \bibinfo {note}
  {http://www.ams.jhu.edu/~eyink/TurbulenceIII/notes.html}\BibitemShut
  {NoStop}%
\end{thebibliography}%

\newpage
\appendix

\section*{Appendix}

\subsection{Sensitivity to filter shape}

\begin{figure}[tbp]
	\includegraphics[trim = 6mm 0mm 0mm 0mm, clip, width=0.49\linewidth]{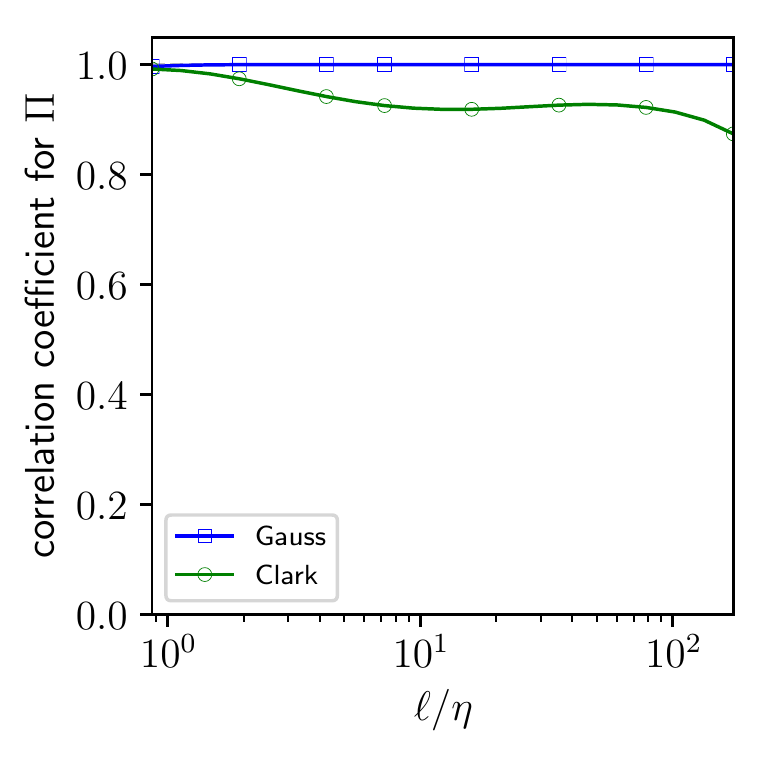}
	\includegraphics[trim = 6mm 0mm 0mm 0mm, clip, width=0.49\linewidth]{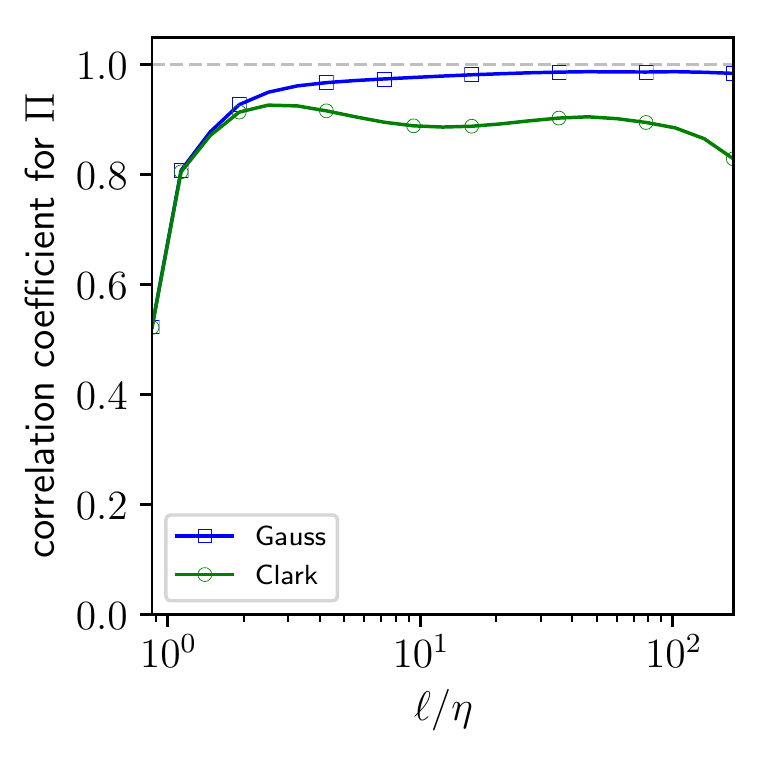}
	\includegraphics[trim = 6mm 0mm 0mm 0mm, clip, width=0.49\linewidth]{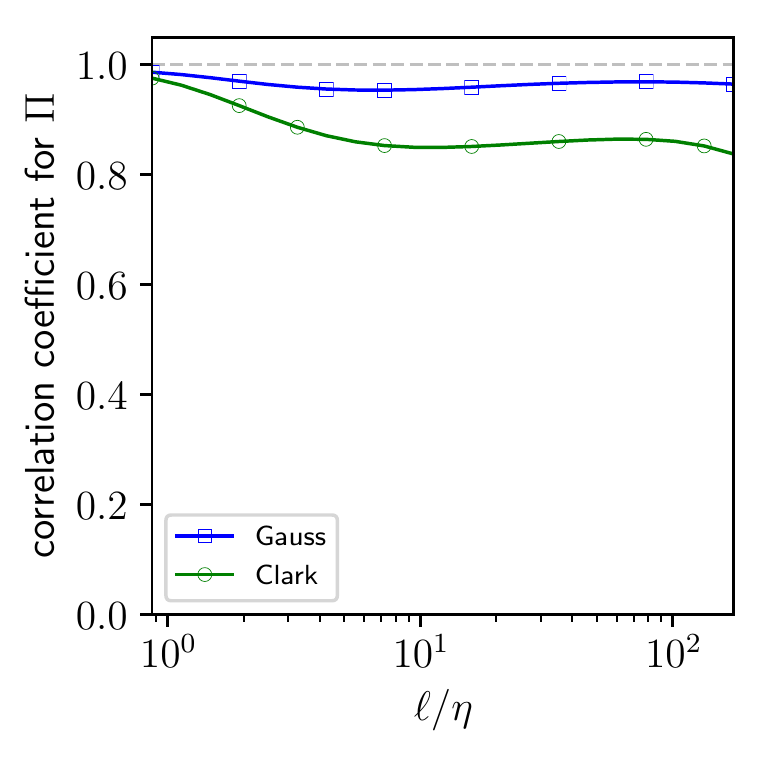}
	\caption{Correlation coefficients for $\Pi$ with Clark model \cite{Clark1979} and with the Gaussian relation, Eq. \eqref{eq:Pi-decomposition} for different filter shapes: (top left) Gaussian filter, (top right) top-hat filter, (bottom) Helmholtz filter \cite{Germano1986}. A gray dashed line indicate a correlation of one.}
	\label{fig:correlation-coefficients}
\end{figure}

The Gaussian filter shape was used extensively in this paper. It is shown here that the results are relatively insensitive to the filter shape and thus reflect the underlying physics of turbulence rather than peculiarities of a particular filter type. Figure \ref{fig:correlation-coefficients} shows correlation coefficients for $\Pi$ computed directly from its definition with $\Pi$ computed using either the Gaussian result (Eq.\ \eqref{eq:Pi-decomposition}) or the nonlinear model of Clark \cite{Clark1979} (first two terms of Eq.\ \eqref{eq:Pi-decomposition}). In the Gaussian filter case (Fig.\ \ref{fig:correlation-coefficients} top), Eq.\ \eqref{eq:Pi-decomposition} is exact and the correlation coefficient is unity for all values of $\ell$. The Clark model still shows a relatively high degree of correlation. For the top-hat filter (Fig.\ \ref{fig:correlation-coefficients} middle), the Clark model shows similar correlation with the exact values. The Gaussian model, with its added scale-nonlocal terms, shows a significant improvement and $\sim 98\%$ correlation for $\ell \gg \eta$. The results for the Helmholtz filter \cite{Germano1986} shown in the bottom panel of Fig.\ \ref{fig:correlation-coefficients} are very similar, with correlation coeffients $\sim 96-97\%$ in the inertial range. The advantage of the Clark model is that it requires no information from scales below $\ell$ and is thus directly applicable to large-eddy simulations (LES). However, if one were to construct an accurate model for the second term in \eqref{eq:stress-split}, a very high fidelity model could be obtained for subgrid stresses in large-eddy simulations.

\begin{figure}[tbp]
	\includegraphics[trim = 0mm 0mm 0mm 0mm, clip, width=0.49\linewidth]{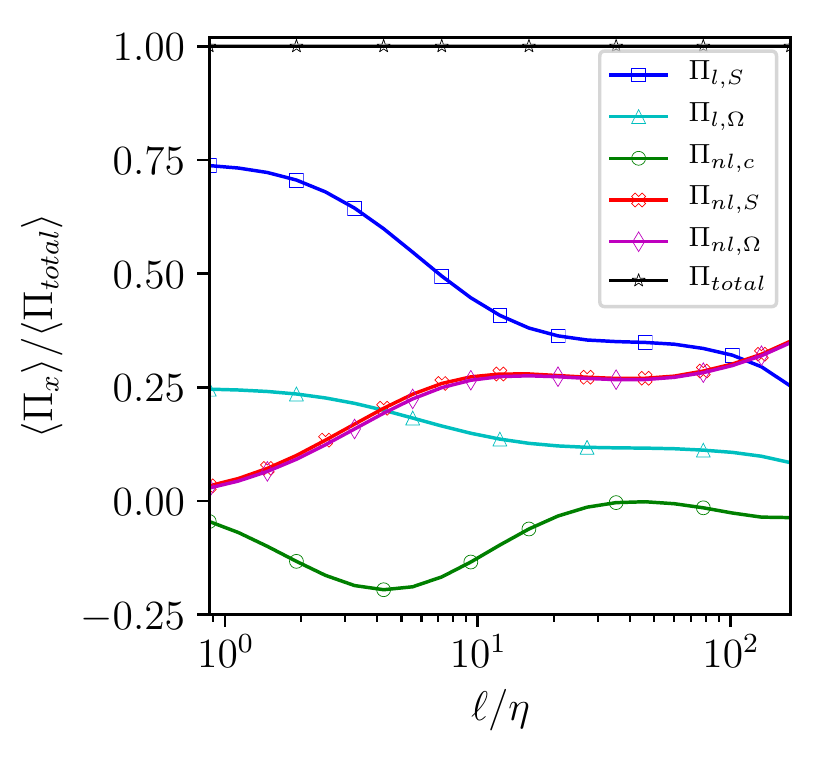}
	\includegraphics[trim = 0mm 0mm 0mm 0mm, clip, width=0.49\linewidth]{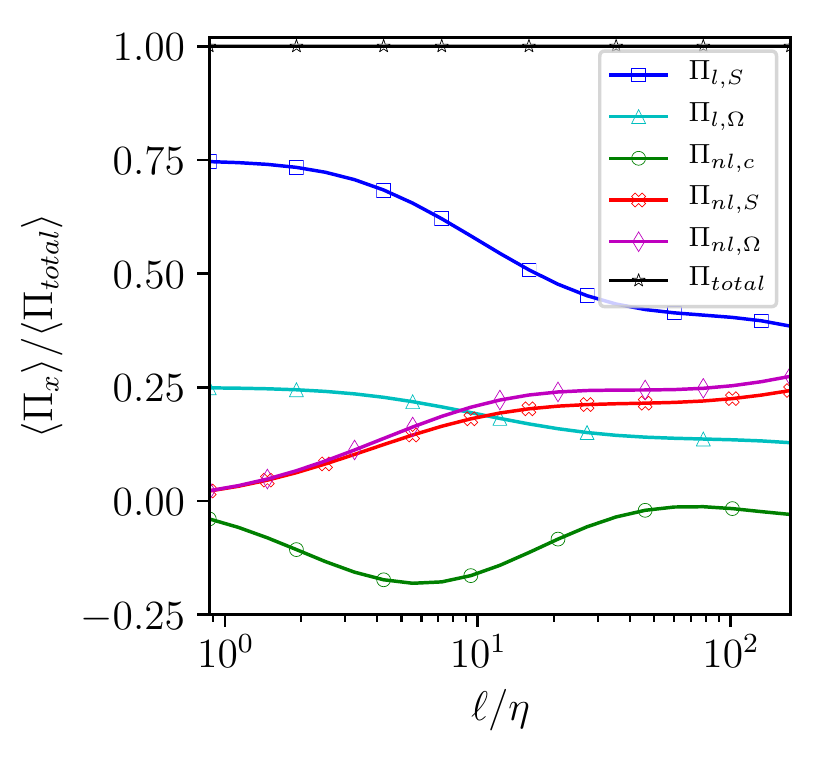}
	\caption{Fraction of net energy transfer accomplished by each of the five mechanisms in Eq. \eqref{eq:Pi-decomposition} for different filter shapes: (left) top-hat filter, (right) Helmholtz filter \cite{Germano1986}.}
	\label{fig:results-sensitivity}
\end{figure}

Now that the applicability of Eq.\ $\eqref{eq:Pi-decomposition}$ has been established for other filter types, the sensitivity of the results shown in this paper is shown in Figure \ref{fig:results-sensitivity}. It is seen that the results for the top-hat and Helmholtz filters are remarkably similar to those of the Gaussian filter shown in Fig.\ \ref{fig:net-transfer}. Some minor discrepancies may be noted. In particular, the Helmholtz filter leads to a larger difference between $\langle \Pi_{nl,S} \rangle$ and $\langle \Pi_{nl,\Omega} \rangle$, but the two remain very close to each other. Also, the scale-local terms are slightly stronger in the Helmholtz filter case. The main conclusions are still applicable for each filter type: both vorticity stretching and strain self-amplification contribute significantly to net inter-scale energy transfer.

This means that the conclusions drawn in this paper using Gaussian filters remain essentially applicable for other filter types. It is worthwhile to mention that the spectral cut-off filter is commonly used for analyzing turbulence. However, this is ill-advised, because the spectral cut-off filter leads to a sub-filter stress tensor which is not positive-definite \cite{Vreman1994b}, which means the sub-filter scale kinetic energy $e^\ell = \tfrac{1}{2} \sigma_{kk}^\ell$ is not guaranteed to be positive. Indeed, negative values of $e^\ell$ have been demonstrated using the spectral cut-off filter  \cite{Vreman1994b}. Therefore, the spectral cut-off filter should not be used in this framework to investigate kinetic energy transfer in turbulence. Instead, such investigations should be limited to non-negative filter kernels, for which sub-filter kinetic energies are provably positive. For the sake of curiosity, the same trends as shown in Figure \ref{fig:net-transfer} and \ref{fig:results-sensitivity} can also be seen in the case of a spectral cut-off filter (not shown), though quantitative deviations are somewhat larger and the correlation is noticeably poorer, see also \cite{Borue1998}.

\subsection{Effect of Reynolds number}

\begin{figure}[tbp]
	\includegraphics{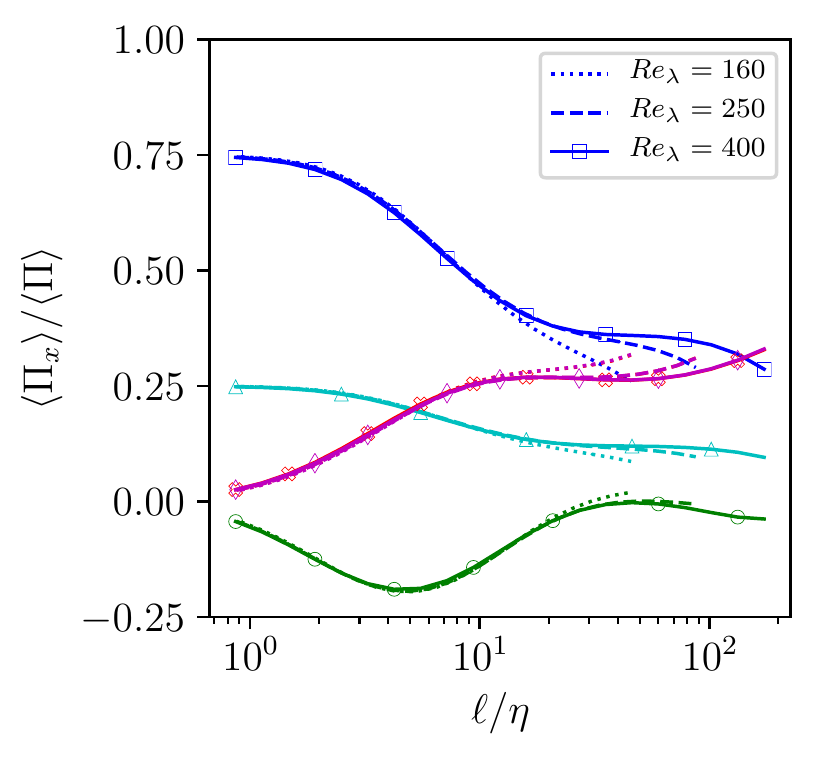}
	\caption{Percent contributions of each term in Eq.\ \eqref{eq:Pi-decomposition} to the net inter-scale energy transfer as a function of filter width $\ell$ for the Gaussian filter. See the caption of Figs.\ \ref{fig:net-transfer} and \ref{fig:results-sensitivity} for the plot type of each mechanism.}
	\label{fig:Re}
\end{figure}

The results presented in this paper were computed from simulations at a relatively modest $Re_\lambda = 400$. It is now shown that the main conclusions should be expected to hold for higher $Re_\lambda$. To this end, simulations at two lower $Re_\lambda$ are considered alongside the $Re_\lambda = 400$ results. The fractional contributions of each term are shown in Figure \ref{fig:Re} for three different Reynolds numbers. The resolution in terms of $k_{\max} \eta \approx 1.4$ is held constant and the grid is refined as $Re_\lambda$ is increased. For $\ell \lesssim 10 \eta$, the curves from all three simulations collapse. Further, the curves from the highest two simulations collapse up to $\ell \lesssim 25 \eta$. As Reynolds number increases. the emergence of the flat regions for each curve (inertial range) is evident. This provides confidence, then, that the results in the investigation can be expected to hold at higher $Re_\lambda$ as the inertial range widens.

\subsection{The extent of nonlocality}

\begin{figure}[tbp]
	\centering
	\includegraphics[width=0.5\linewidth]{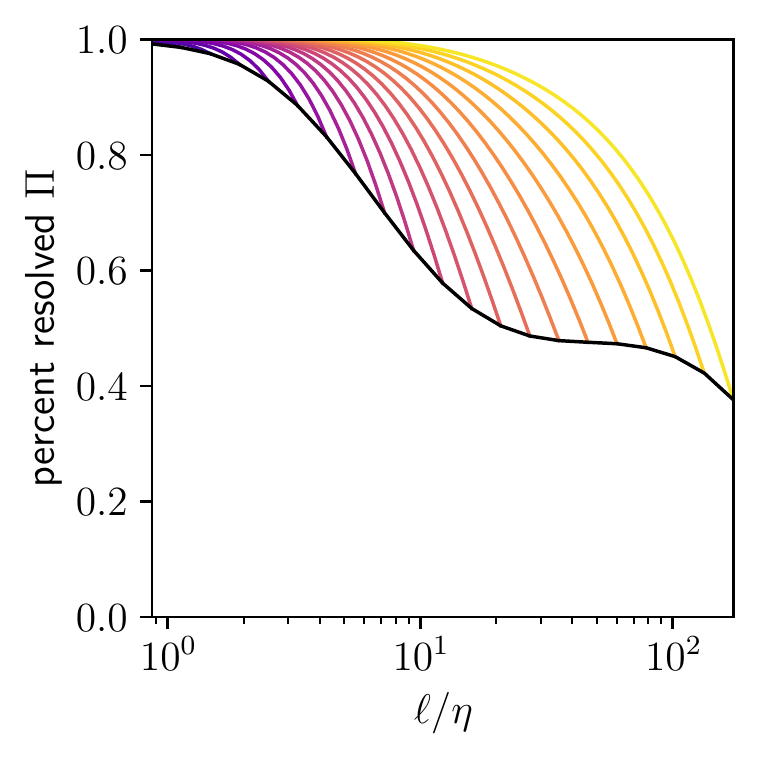}
	\caption{The percent of net energy flux $\langle \Pi^\ell \rangle$ resolved. The lower curve represents resolution of $\Pi^\ell$ at $\ell$ (scale-local) and the curves emanating upward and to the left represent resolution of $\Pi^\ell$ by scales above $\ell^\prime$.}
	\label{fig:transfer-locality}
\end{figure}

The decomposition \eqref{eq:stress-split} identifies strictly scale-local and scale-nonlocal contributions to the energy transfer. Although the idea of the energy cascade in turbulence is quite pervasive, it has long been recognized that the energy transfer is not completely local. The degree to which the scale-nonlocality of energy transfer is dominated by interactions with $\ell^\prime \lesssim \ell$  or $\ell^\prime \ll \ell$ is an important question. This question can be addressed in the present context by generalizing the decomposition \eqref{eq:stress-split},
\begin{multline}
\sigma_{ij}^\ell = \int_{0}^{\ell^2} d\theta ~\overline{\overline{A}_{ik}^{\sqrt{\psi}} \overline{A}_{jk}^{\sqrt{\psi}}}^{\sqrt{\ell^2 - \psi}} \\
+ \int_{0}^{\ell^{\prime 2}} d\theta ~ \overline{\left( \overline{\overline{A}_{ik}^{\sqrt{\theta}} \overline{A}_{jk}^{\sqrt{\theta}}}^{\phi} - \overline{\overline{A}_{ik}^{\sqrt{\theta}}}^{\phi} \overline{\overline{A}_{jk}^{\sqrt{\theta}}}^{\phi} \right)}^{\sqrt{\ell^2 - \ell^{\prime 2}}},
\label{eq:stress-split-general}
\end{multline}
where $\psi = \max\left\lbrace \ell^{\prime 2} , \theta \right\rbrace$ and $\phi = \sqrt{\ell^{\prime 2} - \theta}$. In the limit $\ell^\prime = 0$, the second term vanishes and \eqref{eq:stress} is recovered. In the opposite limit $\ell^\prime = \ell$, the former decomposition, \eqref{eq:stress-split}, into local and nonlocal terms is recovered. For a general $0 \leq \ell^\prime \leq \ell$, this expression separates stresses at scale $\ell$ due to scales above $\ell^\prime$ (the first term) and those below $\ell^\prime$ (the second term). The same steps from \eqref{eq:stress-split} to \eqref{eq:Pi-decomposition} can be followed to generalize the splitting of strain amplification and vorticity stretching decompositions based on the scale $\ell^\prime$. The result is presented in Figure \ref{fig:transfer-locality}. The lower curve is the sum of the two scale-local terms in Figure \ref{fig:net-transfer}, representing the limit $\ell^\prime = \ell$. From each $\ell$ of the lower curve, upward branching curves to the left represent the contribution of the first term in \eqref{eq:stress-split-general} as a function of $\ell^\prime$.

Figure \ref{fig:transfer-locality} demonstrates the relative locality of the energy transfer. The purely local terms, represented by the lower curve, account for roughly half of the net energy transfer when $\ell$ is in the inertial range (see also Figure \ref{fig:net-transfer}). However, including scales down to $\ell^\prime \approx \ell / 4$, the resolution percentage jumps to $\sim 90\%$ or more. In this way, the steep slope of the curves in Figure \ref{fig:transfer-locality} underscore that the nonlocal terms in \eqref{eq:Pi-decomposition} are mostly determined by scales only slightly smaller than $\ell$. This result demonstrates the concept of the `leaky cascade' \cite{Tennekes1972, Lumley1992, Aoyama2005, Eyink2005, Domaradzki2007, Eyink2009, Cardesa2015, Doan2018} in terms of multi-scale vorticity-strain interactions.

\subsection{Spectral blocking and strain self-amplification}

In the introductory section of the main text, it is claimed that the notion of spectral blocking demonstrates the simultaneous necessity of both vorticity stretching and strain self-amplification for sustained transfer of energy from large to small scales. Here, the classical argument of Fjortoft \cite{Fjortoft1953} is briefly recounted fo vorticity stretching, before an anaologous consideration of strain self-amplification is sketched. For simplicity, consider periodic domains such that a Fourier representation exists, following the presentation of \cite{Eyink2014}, see chapter Ib. 

\subsubsection{Enstrophy and spectral blocking}
Consider the volume-integrated enstrophy,
\begin{equation*}
Z(t) = \int \tfrac{1}{2}\omega_i\omega_i d\mathbf{x} = \sum_{\text{all~}k}^{~} \tfrac{1}{2} |\widehat{\omega}(\mathbf{k}, t)|^2 = \sum_{\text{all~}k}^{~} \tfrac{1}{2} |\mathbf{k}|^2 |\widehat{u}(\mathbf{k}, t)|^2,
\end{equation*}
then the incompressible Navier-Stokes equation leads to,
\begin{equation*}
\frac{dZ}{dt} = \int \omega_i S_{ij} \omega_j d\mathbf{x} - \nu \int \frac{\partial \omega_i}{\partial x_j} \frac{\partial \omega_i}{\partial x_j} d\mathbf{x},
\end{equation*}
so that in the absence of vorticity stretching, $\tfrac{dZ}{dt} \leq 0$ and $Z(t) \leq Z(t_0)$ for $t \geq t_0$.

The volume integrated kinetic energy is,
\begin{equation*}
E(t) = \int \tfrac{1}{2} u_i u_i d\mathbf{x} = \sum_{\text{all~}k}^{~} \tfrac{1}{2} |\widehat{u}(\mathbf{k}, t)|^2,
\end{equation*}
and the small-scale kinetic energy, i.e., above a certain wavenumber $\kappa$ is,
\begin{equation*}
E_{>\kappa}(t) = \sum_{|\mathbf{k}| > \kappa}^{~} \tfrac{1}{2} |\widehat{u}(\mathbf{k}, t)|^2.
\end{equation*}
Fjortoft's energy bound may be obtained as follows,
\begin{equation*}
E_{>\kappa}(t) \leq \sum_{|\mathbf{k}| > \kappa}^{~} \frac{|\mathbf{k}|^2}{2 \kappa^2} |\widehat{u}(\mathbf{k}, t)|^2 = \frac{Z_{> \kappa}(t)}{\kappa^2} \leq \frac{Z_{> \kappa}(t_0)}{\kappa^2}.
\end{equation*}
This global (in time) bound on the energy allowed to reach high wavenumbers leads to the notion of spectral blocking preventing the direct energy `cascade' (in the absence of vorticity stretching).

\subsubsection{Dissipation rate}
Consider also the volume-integrated dissipation rate,
\begin{equation*}
\Sigma(t) = \int S_{ij} S_{ij} d\mathbf{x} = \sum_{\text{all~}k}^{~} |\widehat{S}(\mathbf{k}, t)|^2 = \sum_{\text{all~}k}^{~} \tfrac{1}{2} |\mathbf{k}|^2 |\widehat{u}(\mathbf{k}, t)|^2,
\end{equation*}
then the incompressible Navier-Stokes equation leads to,
\begin{equation*}
\frac{d\Sigma}{dt} = - \int 2 S_{ij} S_{jk} S_{ki} d\mathbf{x} - \int \tfrac{1}{2}\omega_i S_{ij} \omega_j d\mathbf{x} - \nu \int \frac{\partial S_{ij}}{\partial x_k} \frac{\partial S_{ij}}{\partial x_k} d\mathbf{x},
\end{equation*}
so that in the absence both vorticity stretching and strain self-amplification, $\tfrac{d\Sigma}{dt} < 0$ and $\Sigma(t) \leq \Sigma(t_0)$.

Note that, like vorticity stretching, strain self-amplification vanishes everywhere in 2D. This is easily seen by considering the two eigenvalues of the strain-rate tensor in 2D, $\lambda_1 + \lambda_2 = 0$, due to incompressibility, so that $S_{ij} S_{jk} S_{ki} = \lambda_1^3 + \lambda_2^3 = 0$.

Removing both vorticity stretching and strain self-amplification, the same derivation above used for enstrophy may be obtained in terms of the dissipation rate. Further, since vorticity stretching appears in the enstrophy and dissipation rate equations with opposite sign, it cannot simultaneously increase both. In other words, to avoid spectral blocking, both vorticity stretching and strain self-amplification are necessary. This is seen by considering the dynamics of both enstrophy and dissipation rate.

In some sense, this is not a particularly interesting exercise, since we have $\Sigma(t) = Z(t)$ and,
\begin{equation*}
-\int S_{ij} S_{jk} S_{ki} d\mathbf{x} = \int \tfrac{3}{4}\omega_i S_{ij} \omega_j d\mathbf{x}
\end{equation*}
from the two Betchov relations \cite{Betchov1956}. Therefore, requiring vorticity stretching for the energy `cascade' is equivalent to requiring strain self-amplification in the scenario considered here. Regardless of how one looks at it, it is clear that the spectral blocking argument does not distinguish between the contributions of vorticity stretching and strain self-amplification in establishing the transfer of energy to small-scales. Rather, it successfully establishes the qualitative insight that both are needed.

\end{document}